\documentclass[11pt,a4paper]{article}
\usepackage{jcappub} % for details on the use of the package, please
\usepackage{bm}% bold math
\usepackage{amsmath}
\usepackage{mathrsfs}
\usepackage[mathscr]{euscript}
\usepackage[compat=1.1.0]{tikz-feynhand}
\usepackage{feynmp}
\usepackage{comment}
\usepackage{natbib}
\usepackage{ulem}
\usepackage[latin1]{inputenc}

\def\mcP{\mathcal{P}}

\def\Mpl{M_{\rm Pl}}

\newcommand{\df}{\text{d}}

\def\0{{(0)}}
\def\sig0{\dot{\sigma}_0}

\def\ph0{\dot{\phi}_0}

\usepackage{braket}

\title{
Inflation with two-form field: the production of primordial black holes and gravitational waves}
\author[a,b]{Tomohiro Fujita,}
\author[c]{Hiromasa Nakatsuka,}
\author[d]{Ippei Obata,}
\author[d,e]{Sam Young}

\affiliation[a]{Waseda Institute for Advanced Study, Waseda University,  Shinjuku, Tokyo 169-8050, Japan}
\affiliation[b]{Research Center for the Early Universe, The University of Tokyo, Bunkyo, Tokyo 113-0033, Japan}
\affiliation[c]{Institute for Cosmic Ray Research, The University of Tokyo, Kashiwa 277-8582, Japan}
\affiliation[d]{Max-Planck-Institut f{\"u}r Astrophysik, Karl-Schwarzschild-Str. 1, 85741 Garching, Germany}
\affiliation[e]{Instituut-Lorentz for Theoretical Physics, Leiden University, 2333 CA Leiden, The Netherlands}

\emailAdd{tomofuji@aoni.waseda.jp,  hiromasa@icrr.u-tokyo.ac.jp, obata@mpa-garching.mpg.de, young@lorentz.leidenuniv.nl}

\abstract{Antisymmetric tensor field (two-form field) is a ubiquitous component in string theory and generally couples to the scalar sector through its kinetic term.
In this paper, we propose a cosmological scenario that the particle production of two-form field, which is triggered by the background motion of the coupled inflaton field, occurs at the intermediate stage of inflation and generates the sizable amount of primordial black holes as dark matter after inflation. We also compute the secondary gravitational waves sourced by the curvature perturbation and show that the resultant power spectra are testable with the future space-based laser interferometers.
}

\begin{document}

\maketitle

%===================================================================%
\section{Introduction}
%===================================================================%

What is the nature of dark matter?
Despite nearly a century of exploration, 
most of its properties are still wrapped in mystery.
In fact, the potential mass scale of dark matter ranges over 90 orders of magnitudes from $10^{-31}~\text{GeV}$ to $10^{60}~\text{GeV}$ \cite{Bertone:2018krk}.
For decades, dark matter searches have focused on particle models for dark matter, especially weakly interacting massive particles (WIMPs).
Recently, however, the non-detection of WIMPs provides strong motivation for the search for alternative dark matter candidates.

Among various dark matter candidates, primordial black holes (PBHs) are well motivated, requiring no extensions to the standard model of particle physics,
and have received a lot of attention owing to the recent detection of gravitational waves (GWs) from merging black hole binaries~\cite{Bird:2016dcv,Clesse:2016vqa,Sasaki:2016jop}.
PBHs can form by the gravitational collapse of local over-dense regions in the early universe \cite{Hawking:1971ei,Carr:1974nx,Carr:1975qj}.
While the mass of PBHs can span many orders of magnitude in principle, the non-detection of PBHs constrains their abundance in many mass ranges.
PBHs which formed with a mass $\lesssim 10^{14}\text{g}$ would have evaporated by today, and could not contribute to the dark matter abundance today.
For light PBHs ($\lesssim 10^{17}\text{g}$), the evaporation of PBHs through Hawking radiation would affect the extragalactic/galactic $\gamma$-ray background \cite{Carr:2009jm,DeRocco:2019fjq,Laha:2019ssq,Laha:2020ivk}.
On the other hand, the abundance of non-evaporating heavy PBHs can be tested by their gravitational effects.
The measurements of gravitational microlensing events limit the abundance of PBHs with mass range around $10^{22}\text{g} - 10^{35}\text{g}$ \cite{Niikura:2017zjd,Allsman:2000kg,Tisserand:2006zx,2011MNRAS.416.2949W}.
Also, stellar-mass PBHs ($\gtrsim10^{33}\text{g}$) would emit high energy photons via gas accretion and could modify the thermal history of universe \cite{Ricotti:2007au,Poulin:2017bwe}.
Hence, PBHs can be the dominant component of dark matter in the intermediate mass window, $10^{17}\text{g} - 10^{22}\text{g}$\footnote{Although there have been other constraints discussed in this mass range such as the femtolensing events of $\gamma$-ray bursts \cite{Barnacka:2012bm}, the dynamical capture of PBHs by stars \cite{Capela:2012jz, Capela:2013yf, Pani:2014rca, Capela:2014ita}, and the ignition of white dwarfs by PBHs \cite{Graham:2015apa}. However, recent studies have revisited these constraints and claimed that this mass window remains open for PBHs as dark matter candidates \cite{Katz:2018zrn, Montero-Camacho:2019jte}.}.

The abundance of PBHs is related to the amplitude of the primordial power spectrum of the curvature perturbation originating from cosmic inflation.
In order for PBHs to compose a significant fraction of dark matter, it is required that a large amplitude of the curvature perturbation is 
produced on scales much smaller than the measurement scale of cosmic microwave background (CMB) anisotropies.
In the simplest class of inflationary models, however, the predicted spectral shape is almost scale-invariant.
From the measurement of CMB anisotropies, the amplitude of curvature perturbations is expected to be too small to predict a sizable amount of PBHs.
To go beyond this naive expectation, many mechanisms for generating large curvature perturbation on small scales have been proposed, including the running mass model \cite{Drees:2011hb}, axion inflation \cite{Lin:2012gs,Bugaev:2013fya,Cheng:2016qzb,Cheng:2018yyr,Ozsoy:2018flq}, a waterfall transition during hybrid 
inflation \cite{GarciaBellido:1996qt,Lyth:2012yp,Bugaev:2011wy,Ashoorioon:2020hln,Spanos:2021hpk,Ashoorioon:2022raz}, a quartic action during inflation and a variable sound speed \cite{Ballesteros:2018wlw}, an inflation coupled with the Gauss-Bonnet term \cite{Kawai:2021edk}, amongst many others.

Among these works, the mechanism of generating curvature perturbations via the particle production of coupled matter sectors has been intensively studied.
Such a representative sector is the form fields. A vector field (dubbed gauge field or one-form field) and an anti-symmetric tensor field (two-form field) are predicted by string theory,
which are naturally coupled to the scalar sector in the low-energy effective action \cite{Lidsey:1999mc,Svrcek:2006yi}.
In this framework, the form field can experience an instability caused by the motion of coupled scalar field and can source the coupled fluctuations on the super-horizon scales during the inflationary epoch, which finally leads to a rich phenomenology such as the generation of PBHs or the accompanied scale-dependent GWs. \cite{Linde:2012bt,Bugaev:2013fya,Domcke:2016bkh,Garcia-Bellido:2016dkw,Garcia-Bellido:2017aan,Domcke:2017fix,Ozsoy:2018flq,Ozsoy:2020kat,Kawasaki:2019hvt, Okano:2020uyr}. 
It is the aim of this paper to explore the particle production of the two-form field and its observable consequences.

In recent years, the phenomenology of particle production of two-form field has also been developed in parallel with studies of gauge field production.
If the generation of the two-form field is on large scales, then a coherent mode of two-form field would break isotropy of universe due to its direction dependence \cite{Ohashi:2013mka,Ohashi:2013qba,Ito:2015sxj}.
It would predict a statistically-anisotropic power spectrum \cite{Obata:2018ilf}, whose spectral anisotropies are different from the case of inflationary model with $U(1)$ gauge field \cite{Fujita:2018zbr}.
On the other hand, the possibility of the generation of two-form field on small scales has been overlooked.
In this paper, we extend our previous work on a $U(1)$ gauge field \cite{Kawasaki:2019hvt} to the particle production of two-form field.
We show that the two-form field can be amplified at an intermediate stage of inflation and can predict a sizable amount of PBHs as dark matter.

This paper is organized as follows.
In Section \ref{2}, we build up the model of the 2-form field kinetically coupled to the inflaton.
In Section \ref{3}, we calculate the curvature power spectrum sourced by two-form field. 
We estimate the PBH abundance in Section \ref{4}.
In Section \ref{5}, we also discuss the generation of tensor perturbations sourced by two-form field and estimate the amount of secondary GWs after inflation.
We finally summarize our study in Section \ref{6}.
Throughout this paper, we will set the units $\hbar = c = 1$ unless otherwise specified.

%===================================================================%
\section{Model Setup}
\label{2}
%===================================================================%

In this paper, we consider the following inflationary model where an antisymmetric two-form field $B_{\mu\nu}$ is kinetically coupled to the inflaton:
\begin{equation}
\mathcal{L} = \dfrac{\Mpl^2}{2}R - \dfrac{1}{2}\partial_\mu\varphi\partial^\mu\varphi - V(\varphi) - \dfrac{1}{12}I(\varphi)^2H_{\mu\nu\rho}H^{\mu\nu\rho} \ ,
\end{equation}
where $H_{\mu\nu\rho} = \partial_\mu B_{\nu\rho} + \partial_\nu B_{\rho\mu} + \partial_\rho B_{\mu\nu}$ is the antisymmetric field strength of the two-form field
and $I(\varphi)$ is a non-trivial kinetic function deriving from the compactified volume of extra dimensions in the framework of string theory \cite{Lidsey:1999mc,Svrcek:2006yi}.
Hereafter, we assume that two-form field induces negligible backreaction to the background inflaton dynamics, which we will confirm in Section ~\ref{3}.
For a gauge condition of two-form field, we can take
\begin{equation}
\partial_i B_{ij}(t,\bm{x}) = 0 \ , \qquad \partial_i B_{0i}(t, \bm{x}) = 0, \label{eq: twoformgauge}
\end{equation}
with the gauge transformation $\delta_g B_{\mu\nu} = \partial_\mu \xi_\nu - \partial_\nu \xi_\mu$ and the redundant degree of freedom $\xi_\mu \rightarrow \xi_\mu + \partial_\mu\chi$.
One can show that $B_{0i}$ is a non-dynamical variable and can be integrated out in the action.
At a quadratic order in perturbation, $B_{0i}$ could couple only to  $B_{ij}$, because $B_{ij}$ is assumed to have no background components. Due to the gauge conditions \eqref{eq: twoformgauge}, however, those quadratic interactions between $B_{0i}$ and $B_{ij}$ vanish and hence $B_{0i}$ gives a higher-order contribution in perturbations.
Therefore, hereafter we neglect $B_{0i}$ in our analysis since its contribution becomes relevant only for the trispectrum.

Throughout this paper, we assume that $B_{\mu\nu}$ does not have its homogeneous component that leads to the breaking of the isotropy of space \cite{Ohashi:2013mka,Ohashi:2013qba,Ito:2015sxj}.
Hence, we consider the spatially-flat FLRW metric, $ds^2 = -dt^2 + a(t)^2d\bm{x}^2 = a(\tau)^2(-d\tau^2 + d\bm{x}^2)$, and the perturbation of two-form field at zeroth-order level: $B_{\mu\nu}(t, \bm{x}) = \delta B_{\mu\nu}(t, \bm{x})$.
We decompose $\delta B_{ij}$ in Fourier space as
\begin{equation}
\delta B_{ij}(t,\bm{x})  = \int\dfrac{d\bm{k}}{(2\pi)^3}\delta B_{k}\epsilon_{ij}(\hat{\bm{k}})e^{i\bm{k}\cdot\bm{x}} \ ,
\end{equation}
where $\epsilon_{ij}$ is the antisymmetric tensor
\begin{equation}
\epsilon_{ij}(\hat{\bm{k}}) = i\left(\begin{array}{ccc}
0 & k_z & -k_y \\
-k_z & 0 & k_x \\
k_y & -k_x & 0
\end{array}\right) = i\left(\begin{array}{ccc}
0 & \cos\theta_{\hat{\bm{k}}} & -\sin\theta_{\hat{\bm{k}}}\sin\varphi_{\hat{\bm{k}}} \\
-\cos\theta_{\hat{\bm{k}}} & 0 & \sin\theta_{\hat{\bm{k}}}\cos\varphi_{\hat{\bm{k}}} \\
\sin\theta_{\hat{\bm{k}}}\sin\varphi_{\hat{\bm{k}}} & -\sin\theta_{\hat{\bm{k}}}\cos\varphi_{\hat{\bm{k}}} & 0
\end{array}\right)
\end{equation}
with $\hat{\bm{k}} = (\sin\theta_{\hat{\bm{k}}}\cos\varphi_{\hat{\bm{k}}}, \ \sin\theta_{\hat{\bm{k}}}\sin\varphi_{\hat{\bm{k}}}, \ \cos\theta_{\hat{\bm{k}}})$ represented in terms of the polar coordinates.
Then, it satisfies the following relationships:
\begin{equation}
k_i\epsilon_{ij}(\hat{\bm{k}}) = 0 \ , \qquad \epsilon_{ij}(-\hat{\bm{k}}) = \epsilon_{ij}^*(\hat{\bm{k}}) \ , \qquad \epsilon_{ij}(\hat{\bm{k}})\epsilon_{ij}^*(\hat{\bm{k}}) = 2 \ .
\end{equation}
Then, the equation of motion (EoM) for $\delta B_{\bm{k}}$ is given by
\begin{equation}
\left[ \partial_x^2 + 1 - \dfrac{\bar{I}_{xx}}{\bar{I}} - \dfrac{2\bar{I}_x}{x\bar{I}} \right]\left(\bar{I}\delta B_{k}/a\right) = 0 \ , \label{eq: EOMB}
\end{equation}
where we ignored the inflaton perturbation $\delta\varphi(t,\bm x)=\varphi(t,\bm x)-\bar{\varphi}(t)$ in the kinetic function $\bar{I}\equiv I(\bar{\varphi}(t))$ because it gives a higher-order contribution in perturbation, and introduced a dimensionless conformal time variable, $x\equiv -k\tau$.
For a varying $\bar{I}$, the EoM is modified and the two-form field can be produced.
To avoid the strong coupling problem, we assume $\bar{I}$ decreases with time.
Here, we define the logarithmic decay rate of the kinetic function
\begin{equation}
n(t) \equiv \dfrac{d\ln\bar{I}}{dN}  
= -\dfrac{\bar{I}_\varphi}{\bar{I}}\dfrac{\dot{\bar{\varphi}}}{H}, \label{eq: n}
\end{equation}
with respect to the number of e-foldings $dN = -H dt$.
As we will see later,  
$\bar{I}$ decreases and the decay rate is positive during inflation, $n(t) > 0$.

To find useful analytic expressions, we first consider the case where $n(t)$ is varying slowly enough to neglect its time-variation: $n\simeq const$. Then \eqref{eq: EOMB} is rewritten as
\begin{equation}
\left[ \partial_x^2 + 1 - \dfrac{n(n+1)}{x^2} \right]\left(\bar{I}\delta B_{k}/a\right) \simeq 0 \ . \label{eq: mode2form}
\end{equation}
With the Bunch-Davies initial condition, the solution is given by the Hankel function of the first kind.
The ``electromagnetic-like" components of the two-from field, $E_{k} \equiv \bar{I}\delta\dot{B}_{k}/a^2$ and $M_k \equiv k\bar{I}\delta B_k/a^3$, are computed as
\begin{align}
E_{k} &= \dfrac{H^2 e^{i\tfrac{n+1}{2}\pi}}{\sqrt{2k^3}}\sqrt{\dfrac{\pi x^5}{2}}H^{(1)}_{n+3/2}(x) \propto x^{1-n} \qquad (x \ll 1) \ , \label{eq: ele} \\
M_{k} &= \dfrac{H^2 e^{i\tfrac{n+1}{2}\pi}}{\sqrt{2k^3}}\sqrt{\dfrac{\pi x^5}{2}}H^{(1)}_{n+1/2}(x) \propto x^{2-n} \qquad (x \ll 1) \ , \label{eq: mag}
\end{align}
where we took the super-horizon limit in right proportional relations.
The contribution to the electric energy density $\rho_E \equiv I^2\dot{B}^2_{ij}/(4a^4)$ from each Fourier mode is proportional to
$|E_{k}|^2$. Thus, the two-form field increases its energy density on the super-horizon scale if $n >1$.
Even if $n(t)$ varies in time and the solution would not be given analytically, the particle production of two-form field occurs on the super-horizon regime for $n\gtrsim 1$ as we will see in Section \ref{3}.
On the other hand, the gradient (magnetic) energy density is suppressed by a factor of $x^2$ in comparison with electric energy density.
Therefore, hereafter we ignore the contribution from the gradient terms.

Regarding the kinetic function, we consider the following functional form \cite{Kawasaki:2019hvt}
\begin{equation}
I(\varphi) = B_1\exp\left(c_1\dfrac{\varphi}{\Mpl}\right) + B_2 \ ,
\label{def kinetic function}
\end{equation}
where $B_{1}, B_{2},$ and $c_1 $ are model parameters which are all positive.
In addition to a conventional exponential function, we also introduce another constant term expected to arise from a string-loop modification in powers of a dilaton-dependent coupling constant \cite{Damour:1994zq}.
We also assume this constant term plays a role of stabilizing the dilatonic field and controlling the generation of two-form field at around the end of inflation.
For the infaton potential, we adopt the Starobinsky model consistent with the current $(n_s, r)$ constraint in CMB observation\footnote{In view of string theory, this kind of potential is known to be arisen from the presence of D-brane defect \cite{Ellis:2014cma}. In any case, however, we emphasis that our prediction is not sensitive to a specific choice of potential.}
\begin{equation}
V(\varphi) = \mu^4(1-e^{-\gamma\varphi})^2 \ , \qquad \gamma = \sqrt{\dfrac{2}{3}}\Mpl^{-1},
    \label{eq:starobinsky_model_potential}
\end{equation}
and solve the background system of inflaton:
\begin{align}
&\ddot{\bar{\varphi}} + 3H\dot{\bar{\varphi}} + V_\varphi = 0 \ , \\
&3\Mpl^2H^2 = \frac{1}{2}\dot{\bar{\varphi}}^2+ V(\bar{\varphi}) \ ,
\end{align}
where we have ignored the backreaction effect from the two-form field and we will discuss the validity of this in the next section.
As discussed in Ref.~\cite{Kawasaki:2019hvt}, one can find approximated analytic expressions
for the e-folds $N(t)$ and the decay rate $n(t)$ under the slow-roll approximation.
However, they do not enable us to compute the power spectrum of the two-form field with sufficient accuracy, and we do not present them here.

\begin{figure}[htbp]
\center
  \includegraphics[width=80mm]{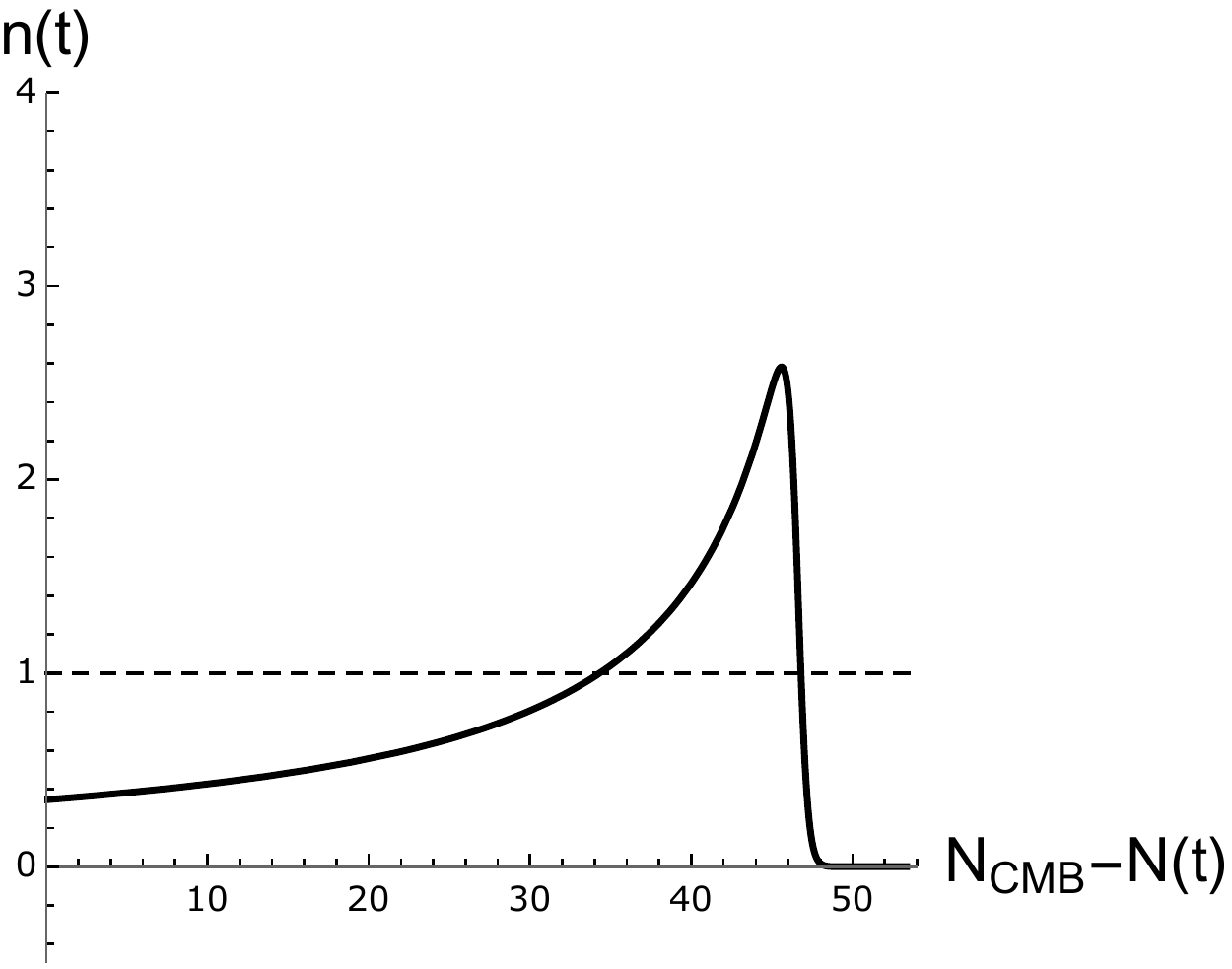}
  \caption{
  The time evolution of $n(t)$ numerically computed with the parameter set \eqref{parameterset}.
  The horizontal axis is the number of e-folds $N_{\rm CMB}-N(t) = \ln (a(t)/a_{\rm CMB})$ and inflation ends at $N_{\rm CMB}-N(t_{\rm end}) \approx 51$.
  $n(t)$ increases roughly inversely proportional to $N$, and 
  it quickly decays a few e-folds before the inflation end.
  The two-form field grows on the super-horizon scale, while $n(t)$ is larger than unity (horizontal dashed line).
  }
 \label{fig:n}
\end{figure}

We set the following values of model parameters\footnote{We don't need to specify the absolute values of $B_{1}$ and $B_2$ because they appear only in the form of ratio $B_2/B_1$ in the equations of motions.}
\begin{equation}
B_2/B_1\simeq 5.5\times10^{16} \ , \qquad c_1 = 15, 
\label{parameterset}
\end{equation}
and numerically solve the background equation for $\varphi(t)$ and 
plot the evolution of the decay rate $n(t)$ in Fig.\ref{fig:n}.
Here, $n(t)$ increases until the background inflaton $\bar{\varphi}(t)$ reaches a value for which the two terms in \eqref{def kinetic function} become comparable. Before that point, the first term in \eqref{def kinetic function} is dominant and the decrease of $\bar{I}$ is accelerated, because the inflaton is accelerated by the potential. After that point, on the other hand,
$\bar{I}$ quickly settles at the constant value $B_2$ and the decay rate $n(t)$ vanishes.
In Fig.\ref{fig:n}, one observes that $n(t)$ is larger than unity for about 15 e-folds, and
the two-form field significantly grows during this period of time.

%===================================================================%
\section{Generation of scalar mode}
\label{3}
%===================================================================%

In this section, we compute the scalar mode induced by the particle production of the two-form field via the interaction Lagrangian,
$\mathcal{L}_{\rm int} = -I(\varphi)^2H_{\mu\nu\rho}H^{\mu\nu\rho}/12$.
At leading order, it is written as
\begin{equation}
\mathcal{L}_{\rm int} \simeq \dfrac{\bar{I}_\varphi\delta\varphi}{2\bar{I}}\dfrac{\bar{I^2}\delta\dot{B}_{ij}^2}{a^4} \ ,
\end{equation}
where we ignored the contribution from the gradient term of two-form field because it is always suppressed on the super-horizon scale  
in comparison with the electric-like term (see \eqref{eq: ele} and \eqref{eq: mag}).
We quantize the fluctuation of the inflaton field and decompose it in Fourier mode as
\begin{equation}
\delta\varphi(t, \bm{x}) = \int\dfrac{d\bm{k}}{(2\pi)^3}\hat{\delta\varphi}_{\bm{k}}e^{i\bm{k}\cdot{\bm{x}}} \ .
\end{equation}
The equation of motion for inflaton in the slow-roll regime is approximately given by
\begin{align}
&\left[\partial_x^2 + 1 - \dfrac{2
}{x^2} \right](a\hat{\delta\varphi}_{\bm{k}}) \simeq a^3\dfrac{2}{k^2}\dfrac{\bar{I}_\varphi}{\bar{I}}\hat{\delta\rho}_{E, \bm{k}} \ , \label{eq: deltavarphi} \\
&\hat{\delta\rho}_{E, \bm{k}} = \dfrac{1}{4}\int\dfrac{d\bm{p}}{(2\pi)^3}\hat{E}_{\bm{p}}\epsilon_{ij}(\hat{\bm{p}})\hat{E}_{\bm{k}-\bm{p}}\epsilon_{ij}(\widehat{\bm{k}-\bm{p}}) \ .
\label{deltarho_E}
\end{align}
%
%///////////////////////////////////////////////////////////////////////////////////%
%///////////////////////////////////////////////////////////////////////////////////%
%
The solution can be split into the vacuum mode and the sourced mode as
\begin{equation}
\hat{\delta\varphi}_{\bm{k}} = \hat{\delta\varphi}_{\bm{k},v} + \hat{\delta\varphi}_{\bm{k},s}\,.
\end{equation}
By using the Green function method, the solution of sourced mode is given by
\begin{align}
a\hat{\delta\varphi}_{\bm{k},s}(x) &= \dfrac{2}{k^2}\int_{-\infty}^{\infty} dy G_R(x,y)a^3(y)\dfrac{\bar{I}_\varphi}{\bar{I}}\hat{\delta\rho}_E \ , \quad y \equiv -k\tau'\,,
\label{Greenfucntion}
\end{align}
with the retarded Green function $G_R(x,y) \equiv -\Theta(y-x)(x^3-y^3)/(3xy)$ obtained by solving the homogeneous solution of \eqref{deltarho_E} in de Sitter approximation.
Since the two inflaton modes are statistically-independent, the power spectrum is written as
\begin{equation}
\langle \hat{\delta\varphi}_{\bm{k}} \hat{\delta\varphi}_{\bm{k'}} \rangle = (2\pi)^3(\bm{k}+\bm{k'})\dfrac{2\pi^2}{k^3}\Big(\mathcal{P}_{\delta\varphi,v}(\bm{k}) + \mathcal{P}_{\delta\varphi,s}(\bm{k})\Big) \ .
\label{powerspectra}
\end{equation}
Without loss of generality, we can choose $\hat{\bm{k}} = (0,0,1)$.
In this case, the following relation of the antisymmetric tensor holds:
\begin{equation}
\epsilon_{ij}(\hat{\bm{p}})\epsilon_{ij}(\widehat{\bm{k}-\bm{p}}) = -2\cos(\theta_{\hat{\bm{p}}} + \theta_{\widehat{\bm{k}-\bm{p}}}) = \dfrac{2}{|\bm{k}-\bm{p}|}(p-k\cos\theta_{\hat{\bm{p}}}) \ .
\end{equation}
Combining \eqref{deltarho_E}, \eqref{Greenfucntion} and \eqref{powerspectra}, we obtain
\begin{align}
\mathcal{P}_{\delta\varphi,s}(\bm{k}) = \dfrac{k^3}{\pi^2H^4}\int\dfrac{d\bm{p}}{(2\pi)^3}\cos^2(\theta_{\hat{\bm{p}}} + \theta_{\widehat{\bm{k}-\bm{p}}})\left[\int^\tau_{\tau_{\text{min}}}\dfrac{d\tau'}{\tau'}\dfrac{\bar{I}_\varphi}{\bar{I}}\dfrac{y^3-x^3}{3y^3}E_pE_{|\bm{k}-\bm{p}|} \right]^2 \ , \label{eq: Ps}
\end{align}
where we approximate the lower bound of the time integral by $|\tau_{\text{min}}| = \text{min}(1/p, 1/|\bm{k}-\bm{p}|)$, because the two-form field can grow only after the horizon crossing and we focus on the contribution from the super-horizon modes. 

\begin{figure}[tbp]
\center
  \includegraphics[width=120mm]{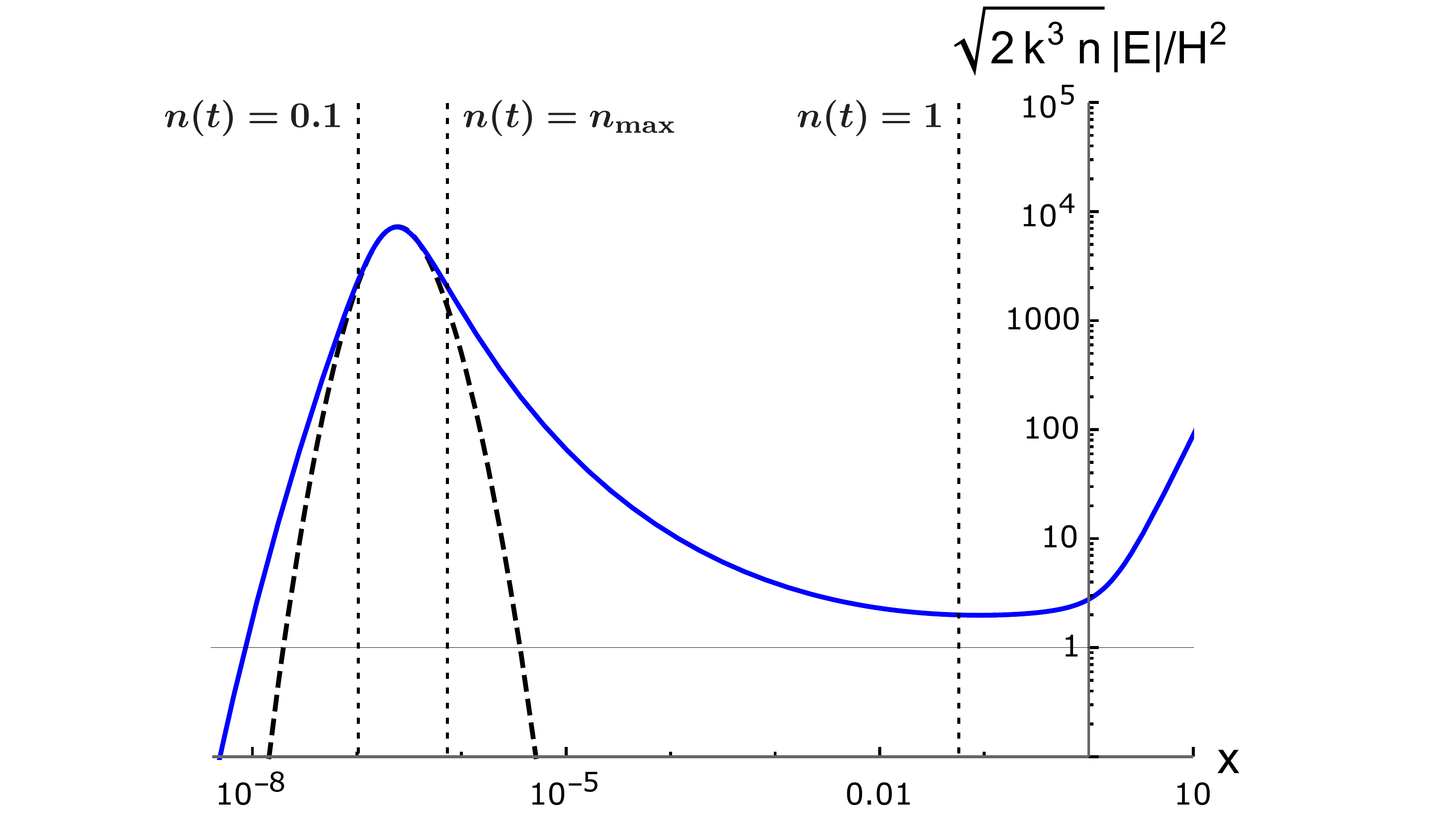}
  \caption
 {The time evolution of $E_k$ which crosses the horizon at 25 e-folds before the inflation end. 
 We normalize it to be dimensionless and multiply $\sqrt{n} \propto \sqrt{I_\varphi/I}$ to match the source term of the inflaton perturbation in \eqref{eq: deltavarphi}.
 The solid blue line shows the exact numerical solution of \eqref{eq: EOMB}.
 The dashed black line shows the fitting function \eqref{eq: fitting} with $\sigma^2 = 1.0$ and $|\tau_{\rm peak}| \sim 2\times10^{-18} \ \text{Mpc}$.
 The vertical dotted lines denote the time when $n(t)=1, n_\mathrm{max}$ and $0.1$ from right to left.
}
 \label{fig:E}
\end{figure}

\begin{figure}[tbp]
\center
  \includegraphics[width=100mm]{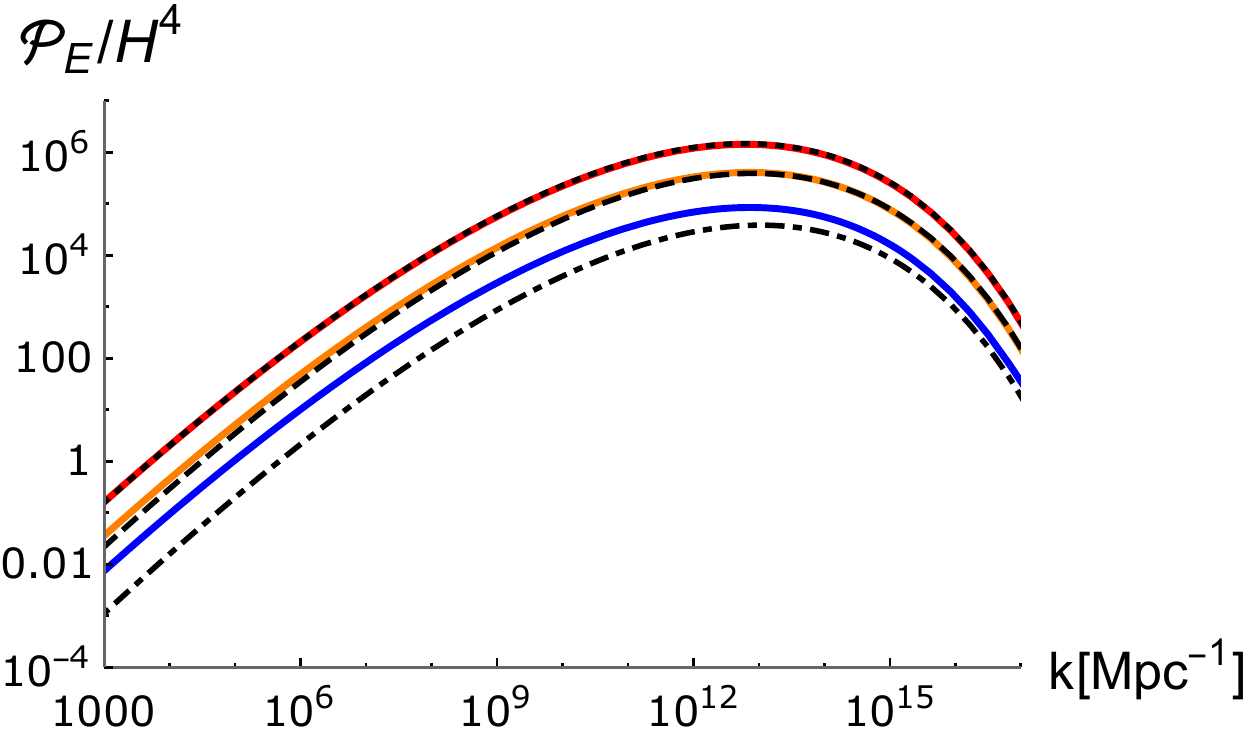}
  \caption
 {
The power spectrum of the two-form field $\mathcal{P}_E(k)$ at several e-folds; $N=N(t_1)\simeq 4$ (solid red), $N=N(t_1)-0.5$ (solid orange) and $N=N(t_1)-1$ (solid blue), where $t_1$ denotes the time when the growth of the two-form field stops, $n(t_1)=1$, and the spectral peak gets a maximum value. The black lines show our fitting function \eqref{eq: fitting} evaluated at $N=N(t_1)$ (dotted) and $N=N(t_1)-0.5$ (dashed), and $N=N(t_1)-1$ (dot-dashed). They agree well until the peak amplitude becomes around two orders of magnitude less than that at $t = t_1$.
 }
 \label{fig:rhoE}
\end{figure}

To perform the time and momentum integrals in \eqref{eq: Ps} separately, we find an analytical formula for $E_k(\tau)$.
The electric mode function of the two-form field is well approximated 
by the following Gaussian fitting function on the scales where the particle production is relevant:
\begin{equation}
E_k \simeq \dfrac{H^2}{\sqrt{2k^3}}E_{\text{peak}}(k)\exp\left[ -\dfrac{(\ln(\tau/\tau_{\text{peak}}))^2}{\sigma^2} \right] \ , \label{eq: fitting}
\end{equation}
where $E_{\text{peak}}(k)$ is a maximum amplitude of $E_k$ at $\tau=\tau_{\rm peak}$ 
and $\sigma$ characterizes the time period for which $E_k$ stays around the peak amplitude.
We simply use the numerical result for $E_{\rm peak}(k)$.
$\tau_{\rm peak}$ corresponds to the time when $n(t)$ falls below unity at $N_{\rm CMB}-N\approx 49$
where the growth of the two-form field stops.
$\sigma$ is determined by $n(t)$ which is governed by the background dynamics, and hence $\sigma$ does not depend on $k$. 
Fig.~\ref{fig:E} shows the time evolution of the electric mode function crossing the horizon at $N_k\approx 25$
, and compare it with the fitting solution.
One can see that \eqref{eq: fitting} fits well the numerical solution in the time domain.
We also plot the time evolution of the dimensionless power spectrum of the electric field $\mathcal{P}_E(k) \equiv k^3|E_k|^2/(2\pi^2)$ in Fig.~\ref{fig:rhoE} and find that this fitting function agrees well with all the relevant modes at around the time when the particle production occurs significantly.
%
%///////////////////////////////////////////////////////////////////////////////////%
%///////////////////////////////////////////////////////////////////////////////////%
%

Using \eqref{eq: fitting}, we rewrite \eqref{eq: Ps} in the late time limit as
\begin{align}
\mathcal{P}_{\delta\varphi,s}(\bm{k})|_{\tau\rightarrow0} &\simeq \dfrac{H^4\mathcal{F}^2}{72\Mpl^2\pi^2\epsilon_H}\int\dfrac{d\bm{p}^*}{(2\pi)^3}\dfrac{\cos^2(\theta_{\hat{\bm{p}}} + \theta_{\widehat{\bm{k}-\bm{p}}})}{p^{*3}|\bm{k}-\bm{p}|^{*3}}E^2_{\text{peak}}(p)E^2_{\text{peak}}(|\bm{k}-\bm{p}|) \ , \label{eq: Ps2} \\
\mathcal{F} &\equiv \int_{-\infty}^\infty \dfrac{d\tau'}{\tau'}~n(\tau')\exp\left[-2\dfrac{\ln (\tau'/\tau_{\text{peak}})^2}{\sigma^2}\right] \label{eq: F} \ ,
\end{align}
where $
\bm{p}^* \equiv \bm{p}/k \ , \ |\bm{k}-\bm{p}|^* \equiv |\bm{k}-\bm{p}|/k$ and we have used the slow-roll approximation $\bar{I}_\varphi/\bar{I} \simeq n/(\Mpl\sqrt{2\epsilon_H})$.
We notice that the time integration of $\mathcal{F}$ can be extended to infinity because its integral receives its support almost around the peak $\tau = \tau_{\rm peak}$.
We numerically integrate the time and momentum integrals in \eqref{eq: Ps2} and obtain $\mathcal{P}_{\delta\varphi,s}$.

\begin{figure}[tbp]
\center
  \includegraphics[width=100mm]{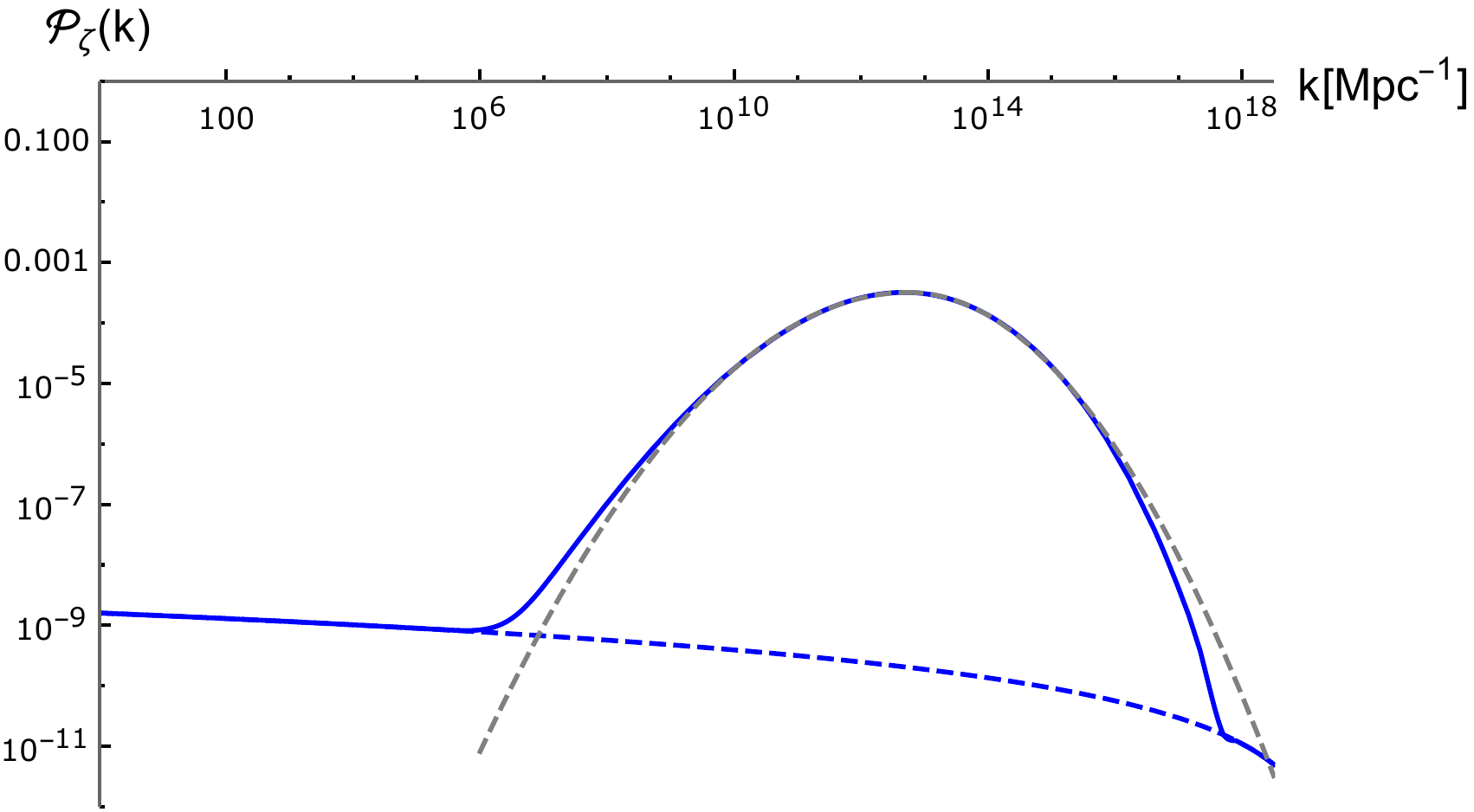}
  \caption
 {Power spectrum of the curvature perturbation $\mathcal{P}_{\zeta} = \mathcal{P}_{\zeta, v} + \mathcal{P}_{\zeta,s}$  against the wave number $k [\text{Mpc}^{-1}]$.
 The solid blue line represents the full result which numerically estimates \eqref{eq: Ps2}, whereas the dashed blue line shows the contribution only from the vacuum mode.
 The black dashed line denotes the approximate fitting function \eqref{eq: gaus} with the parameter set \eqref{eq: paraset}.
 }
 \label{fig:zeta}
\end{figure}

Now we convert the obtained $\mathcal{P}_{\delta\varphi,s}$ into the power spectrum of the curvature perturbation using the relation on the flat-slicing, $\zeta = -H\delta\varphi/\dot{\bar{\varphi}}$.
Corresponding to \eqref{powerspectra}, the curvature power spectrum is also separated into two contributions
\begin{equation}
\mathcal{P}_{\zeta}(k) = \mathcal{P}_{\zeta, v}(k) + \mathcal{P}_{\zeta,s}(k) \ .
\end{equation}
The first term $\mathcal{P}_{\zeta,v} = H^2/(8\pi^2\Mpl^2\epsilon_H)$ is the power spectrum of the vacuum mode normalized as $\mathcal{P}_{\zeta,v}(k_{\text{CMB}}) \simeq 2.1\times10^{-9}$ on CMB scale.
On the other hand, the second term denotes that of the sourced mode which has a peak around $k=k_p \simeq 10^{13}\text{Mpc}^{-1}$, inheriting a similar feature as that seen in $\mcP_E(k)$.
Around its peak, the numerically computed $\mathcal{P}_{\zeta,s}(k)$ is well described by the following fitting function:
\begin{equation}
\mathcal{P}_{\zeta, s}(k) \simeq A \exp\left[-\dfrac{(\ln(k/k_p))^2}{\sigma_\zeta^2}\right] \ . \label{eq: gaus}
\end{equation}
In the plot of Fig.~\ref{fig:zeta}, we set
\begin{equation}
A \simeq 3.2\times10^{-4} \ , \quad k_p \simeq 5.6\times 10^{12}\text{Mpc}^{-1} \ , \quad \sigma^2_\zeta \simeq 3.7^2\Theta(k_p-k) + 3.1^2\Theta(k - k_p). 
\label{eq: paraset}
\end{equation}
In Fig.~\ref{fig:zeta}, one can confirm the agreement between the numerical and the fitting results. We will also use the same values of the parameters as \eqref{eq: paraset} to estimate the abundance of PBHs in the next section. 
%
%///////////////////////////////////////////////////////////////////////////////////%
%///////////////////////////////////////////////////////////////////////////////////%
%
Before closing this section, we estimate the backreaction effect of the two-form field on our background dynamics. The electric energy density of the two-form field
\begin{equation}
\langle \rho_E \rangle = \int d\ln p\dfrac{p^3}{4\pi^2}|E_p|^2 = \dfrac{1}{2}\int d\ln p ~\mathcal{P}_E(p)
\end{equation}
is included in the Friedmann equation and the equation of motion for inflaton as
\begin{align}
&3\Mpl^2H^2 = \frac{1}{2}\dot{\varphi}^2+ V(\varphi) + \langle \rho_E \rangle \ , \\
&\ddot{\bar{\varphi}} + 3H\dot{\bar{\varphi}} + V_\varphi = \dfrac{2\bar{I}_\varphi}{\bar{I}}\langle \rho_E \rangle \ .
\end{align}
We can roughly estimate the magnitude of $\langle \rho_E \rangle$ as
\begin{equation}
\langle \rho_E \rangle \sim \dfrac{1}{2}\Delta N \mathcal{P}_E(k_{\rm peak}) \ ,
\end{equation}
where $\Delta N = \mathcal{O}(1)$ is the logarithmic width of the power spectrum at around the peak $k=k_{\rm peak}$.
Therefore, the backreaction of the energy density of two-form field is negligible when the following conditions holds:
\begin{align}
&\dfrac{\langle \rho_E \rangle}{3\Mpl^2H^2} \ll 1 \quad \leftrightarrow \quad \dfrac{\mathcal{P}_E(k_{\rm peak})}{H^4} \ll \dfrac{6\Mpl^2}{H^2\Delta N} \sim \dfrac{12}{\pi^2r_v\mathcal{P}_{\zeta, v}\Delta N} \ , \label{eq: back1} \\
&\dfrac{2\langle \rho_E \rangle}{3\Lambda H\dot{\varphi}} \ll 1 \quad \leftrightarrow \quad \dfrac{\mathcal{P}_E(k_{\rm peak})}{H^4} \ll \dfrac{6\Mpl^2\epsilon_H}{H^2n_{\rm max}\Delta N} \sim \dfrac{3}{4\pi^2n_{\rm max}\mathcal{P}_{\zeta, v}\Delta N} \ , \label{eq: back2}
\end{align}
where $r_v$ is a tensor-to-scalar ratio of vacuum modes.
Therefore, \eqref{eq: back1} is automatically satisfied if \eqref{eq: back2} holds.
The amount of r.h.s. in \eqref{eq: back2} is roughly $10^7$ around the spectral peak, which is greater than l.h.s. by an order of magnitude (see Figure \ref{fig:rhoE}).
Thus, we can safely ignore the backreaction of the two-form field at the background level.

%
%///////////////////////////////////////////////////////////////////////////////////%
%///////////////////////////////////////////////////////////////////////////////////%
%

%===================================================================%
\section{Estimate of primordial black hole mass function}
\label{4}
%===================================================================%

In this section we will discuss the generation of PBHs in our model. We will consider the formation of PBHs from the collapse of large-amplitude scalar perturbations during the radiation dominated epoch. When a perturbation, which is initially super-horizon re-enters the horizon, it can quickly collapse to form a PBH if above some threshold amplitude. In order to estimate the abundance, we use the Press-Schechter-type (PS) formalism \cite{1974ApJ...187..425P}. A more accurate constraint on the power spectrum can be obtained by using peaks theory \cite{Bardeen:1985tr} (as well as modifications to peaks theory for application to PBHs \cite{Young:2020xmk,Germani:2019zez}). However, using the Press-Schechter formalism is straightforwards to adapt to non-Gaussian distributions - which needs to be accounted for since the abundance of PBHs depends strongly on any non-Gaussianity \cite{Bullock:1996at,Ivanov:1997ia,PinaAvelino:2005rm,Lyth:2012yp,Shandera:2012ke,Bugaev:2013vba,Byrnes:2012yx,Young:2013oia,Young:2014oea,Young:2015kda,Young:2015cyn}. The PS approach provides constraints on the power spectrum which are accurate to $\mathcal{O}(10\%)$ when compared to peaks theory calculations \cite{Gow:2020bzo}, which is sufficient for our purposes here. 

If the amplitude of a density perturbation is large enough, it will collapse to form a PBH upon horizon entry. In order to determine whether a perturbation will collapse, the appropriate criterion to use is the compaction function \cite{Young:2014ana,Musco:2018rwt,Young:2019osy} defined as
\begin{equation}
C(\mathbf{x},r) \equiv 2 \frac{\delta M(\mathbf{x},r,t)}{R(t,r)},
\end{equation}
where $\delta M(\mathbf{x},R)$ is the mass excess within a sphere of areal radius $R$ centred at location $\mathbf{x}$, where we have used $G=1$ in this section for convenience. We note that this is equivalent to the smoothed density contrast evaluated at horizon entry if the transfer function is neglected. On super-horizon scales, the time-dependence of $\delta M$ and $R$ cancel, making $C$ a time-independent quantity. On super-horizon scales, at the centre of spherically symmetric peaks, the compaction function is related to the curvature perturbation $\zeta$ as \cite{Musco:2018rwt}
\begin{equation}
C(\mathbf{x},r) = -\frac{2}{3}r \zeta'(r)\left( 2 + r \zeta'(r) \right), 
\end{equation}
where the prime denotes a spatial derivative (which here only includes the radial term due to the spherical symmetry). We have assumed spherical symmetry because the large, rare peaks, from which PBHs form, are expected to be close to spherically symmetric \cite{Bardeen:1985tr}. The notation $\zeta'(r)$ then represents the value of $\zeta'$ at a distance $r$ from the centre of a peak located at $\mathbf{x}$.

Typically, $\zeta$ is expected to follow a Gaussian distribution. However, in the model discussed here, the large contribution to the power spectrum $\mathcal{P}_{\zeta,s}$ is highly non-Gaussian and is expected to follow a $\chi$-squared distribution, allowing us to write:
\begin{equation}
\zeta = \chi^2,
\end{equation}
where $\chi$ is a Gaussian variable - which will be helpful because the derivative of a Gaussian-distributed variable remains Gaussian. The compaction is then related to $\chi$ as
\begin{equation}
C(\mathbf{x},r) = -\frac{4}{3}r \chi'(r)\chi(r)\left( 2 + 2 r \chi'(r)\chi(r) \right),
\end{equation}
where the component $-r \chi'(r)$ is what one would expect from smoothing the second derivative of $\frac{1}{3} \chi$ with a top-hat smoothing function of radius as below $r$ \cite{Young:2019yug}. 
\begin{equation}
\chi_{2,r} = -r \chi'(r) = \frac{r^2}{3} \int \mathrm{d}^3 \mathbf{y} \chi''(\mathbf{x}-\mathbf{y})W(\mathbf{y},r), 
\end{equation}
where the subscript $r$ indicates a smoothing at scale $r$, and the subscript $2$ denotes the second spatial derivative. The factor of $r^2$ is included as the compaction is calculated as $1/r$, whilst the smoothing function (below) includes a term $1/r^3$. The top hat smoothing function $W(\mathbf{x},r)$ is given by
\begin{equation}
W(\mathbf{x},r) = \frac{\Theta \left( r - \left| x \right| \right)}{\frac{4}{3}\pi r^3},
\end{equation}
where $\Theta$ is the Heaviside step function. The Fourier transform of the smoothing function is
\begin{equation}
    \tilde{W}(k,r) = 3\frac{\sin(k r)-kr\cos(kr)}{(kr)^3}.
\end{equation}

In principle, $\chi(r)$ and $\chi_{2,r}$ are correlated Gaussian variables, and their respective PDFs should be integrated over to find the PDF of $C(\mathbf{x},r)$. However, in the high peak limit relevant for PBHs, the PDF of $\chi(r)$ given a value of $\chi_{2,r}$ is well approximated by a Dirac-delta function. For typical PBH-forming perturbations, we can make the substitution $\chi(r_m) \approx -0.756 \chi_{2,r}$ without significant loss in accuracy, where $r_m$ is the smoothing scale at which the compaction peaks \cite{Young:2022phe} (see also appendix \ref{app:profiles} for a brief discussion). This allows us to write the compaction as a function of a single variable,
\begin{align}
C(\mathbf{x},r) \approx 0.336 \chi_{2,r}^2\left( 2 - 0.504 \chi_{2,r}^2 \right) \notag \\
 \approx 0.336 C_1\left( 2 - 0.504 C_1 \right),
 \label{eqn:compaction}
\end{align}
where in the second equality, for convenience we have rewritten $\chi_{2,r}^2=C_1$ (to represent the linear component of the compaction). We here note that, since this follows a quadratic relationship, there is a maximum value for the compaction $C_\mathrm{max}=2/3$ at $C_{1,\mathrm{max}}=1.982$. Perturbations with $C_1<C_{1,\mathrm{max}}$ are referred to as type I perturbations, and perturbations with $C_1>C_{1,\mathrm{max}}$ are referred to as type II perturbations \cite{Musco:2018rwt}. We will not consider further type II perturbations for 2 reasons: firstly, because the abundance of such perturbations is exponentially suppressed and has a negligible impact on PBH abundance; and secondly, the evolution of type II perturbations cannot be simulated simply using density perturbations and is thus not well understood.

\begin{figure}[tbp]
\center
  \includegraphics[width=100mm]{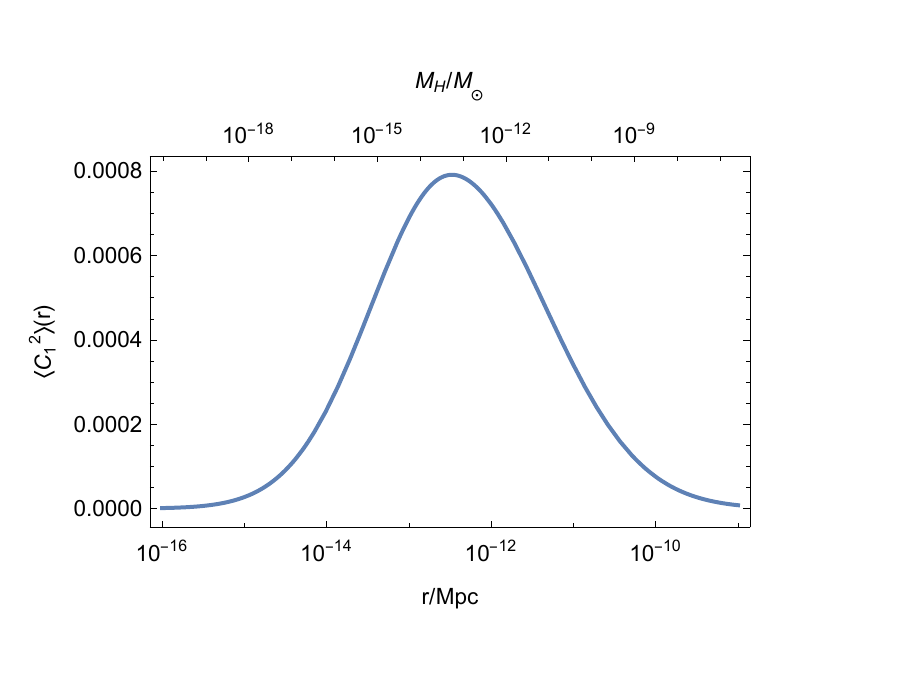}
  \caption
 {
 The variance of the linear component of the compaction, $\langle C_1^2 \rangle$, is plotted against the smoothing scale $r$, for the power spectrum given in equation \eqref{eq: gaus}. PBH formation peaks strongly at the scale where the variance of perturbations is largest.
 }
 \label{fig:variance}
\end{figure}

$C_1$ follows a $\chi$-squared distribution given by
\begin{equation}
    P\left(C_1\right) = \frac{1}{\sqrt{2\pi\sigma^2 C_1}}\exp\left( -\frac{C_1}{2\sigma^2} \right).
\end{equation}
The variance is calculated from the power spectrum $\mathcal{P}_{\zeta,s}$ as a function of the smoothing scale $r$ as
\begin{equation}
2\sigma^4(r)=\langle C_1^2 \rangle(r) = \int\limits_0^\infty \frac{\mathrm{d}k}{k}(kr)^4 \mathcal{P}_{\zeta,s}(k)\tilde{W}^4(k,r).
\end{equation}
The variance $2\sigma^2$ is plotted in Fig. \ref{fig:variance} as a function of $r$, for the power spectrum given in equation \eqref{eq: gaus}, with parameter choices as in equation \eqref{eq: paraset}.

In order to determine the abundance of PBHs, we need to know the mass of PBH which will form from a perturbation, and this depends on both the scale $r$ and amplitude $C$ of the perturbation as \cite{Musco:2008hv,Musco:2012au,Young:2019yug}
\begin{equation}
    M_{\mathrm{PBH}}(C,r) = K M_\mathrm{H}(r)\left(C-C_c\right)^\gamma,
\end{equation}
where $M_\mathrm{H}(r)$ is the horizon mass of the unperturbed background when the horizon scale is equal to the perturbation scale, $(aH)^{-1}=r$, $C_c\approx0.5$ is the critical value for PBH formation (with corresponding value for the linear component $C_{1,c}\approx0.992$). The value of $K$ depends mildly on the specific profile shape of the perturbation collapsing, but we will here simply take $K=4$, which is valid for typical profile shapes, and $\gamma=0.36$ during radiation domination  \cite{Young:2019yug}. 

Substituting equation \eqref{eqn:compaction} to write the PBH mass in terms of $C_1$, we obtain
\begin{equation}
    M_{\mathrm{PBH}}(C_1,r) = K M_\mathrm{H}(r)\left(0.336 C_1\left( 2 - 0.504 C_1 \right)-C_c\right)^\gamma.
\end{equation}
Whilst there is no upper limit for the horizon mass which can form a PBH of given mass, there is a lower limit, given by 
\begin{equation}
    M_\mathrm{min} = \frac{1}{\left( 2/3 - C_c \right)^{\gamma} K} M_\mathrm{PBH} \approx 0.477 M_\mathrm{PBH}.
\end{equation}
Solving to write $C_1$ as a function of the mass, which will be useful later, gives
\begin{equation}
    C_1(M_\mathrm{PBH}M_\mathrm{H}) = \frac{0.336-\sqrt{0.336\left( 0.336-0.504\left( C_c+\left(\frac{M_\mathrm{PBH}}{K M_\mathrm{H}}\right)^{1/\gamma} \right) \right)}} {0.170},
    \label{eqn:CofM}
\end{equation}
where we have kept only the solution corresponding to type 1 perturbations.

%
%///////////////////////////////////////////////////////////////////////////////////%
%///////////////////////////////////////////////////////////////////////////////////%
%

The horizon mass associated with a particular smoothing scale $r$ is given by \cite{Gow:2020bzo}
\begin{equation}
    M_\mathrm{H}(r) \approx 0.5 \left( \frac{g_*}{10.75} \right)^{-1/6}\left( \frac{r}{10^{-6} \rm Mpc} \right)^2 M_\odot,
\end{equation}
where $g_*$ is the number of relativistic degrees of freedom (changes in $g_*$ have a minimal effect due to the power of $1/6$ and will be neglected here). This allows us to relate the smoothing scale $r$ to a specific horizon mass $M_\mathrm{H}$ - and therefore calculate the variance as a function of horizon mass, $2\sigma^2=2\sigma^2(M_\mathrm{H})$.

The fraction of the universe collapsing to form PBHs at the time of formation (we consider this as the time of horizon entry \cite{Young:2019osy}) can be calculated by following a PS formalism by integrating over the range of perturbations which form PBHs:
\begin{equation}
    \beta(M_\mathrm{H}) = 2 \int\limits_{C_{1,c}}^{C_{1,max}}\mathrm{d}C_1 \frac{M_\mathrm{PBH}}{M_\mathrm{H}}P\left(C_1,M_\mathrm{H}\right),
    \label{eqn:beta}
\end{equation}
where the scale-dependence is encoded in the horizon mass $M_\mathrm{H}$. 
Assuming that the universe evolves strictly as radiation-dominated up until the time of matter-radiation equality, we can calculate the density parameter for PBHs at matter-radiation equality:
\begin{equation}
    \left. \Omega_\mathrm{PBH}\right|_\mathrm{eq} = \int \frac{\mathrm{d}M_\mathrm{H}}{M_\mathrm{H}}\left( \frac{M_\mathrm{eq}}{M_\mathrm{H}} \right)^{1/2}\beta\left( M_\mathrm{H} \right),
\end{equation}
where $M_\mathrm{eq}=2.8\times 10^{17}M_\odot$ is the horizon mass at matter-radiation equality (using the same parameter choices as \cite{Nakama:2016gzw}). The term $\left( M_\mathrm{eq}/M_\mathrm{H} \right)^{1/2}$ is included to account for the redshift of PBH energy density (which evolves as matter) during radiation domination. Solving the integral numerically, we find that PBHs form with the correct abundance to make up the entirety of dark matter when the amplitude of the power spectrum is $A\simeq 3.2\times 10^{-4}$, for the power spectrum given in equation \eqref{eq: gaus}.

\begin{figure}[thbp]
\begin{center}
\includegraphics[width=0.8\textwidth]{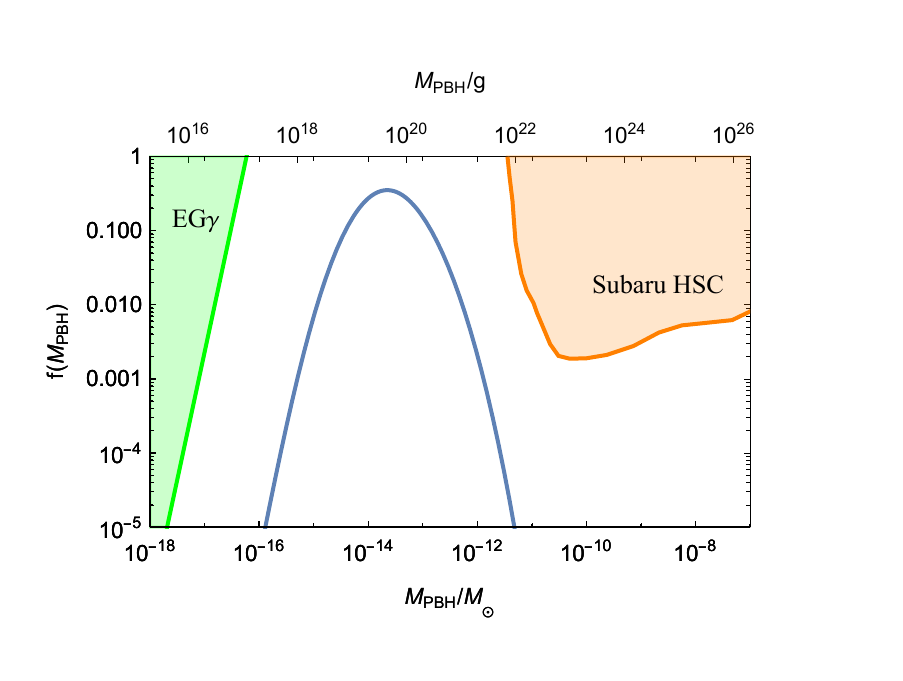}
\end{center}
\caption{
The mass spectrum of PBHs in our model when we choose the parameter set in \eqref{parameterset}.
The green region around $M<\mathcal O(10^{17}) \text g$ is excluded by the extragalactic gamma ray \cite{Carr:2009jm} and the orange region  $\mathcal O(10^{22}) \text g <M $ is excluded by the Subaru/HSC \cite{Niikura:2017zjd}. We note that the constraints plotted are the constraints on a monochromatic PBH mass function (whilst the black line represents the extended mass function predicted by the model presented here) - and so are not completely applicable to our mass function. However, the constraints for extended mass functions are similar to the constraints for monochromatic mass functions. 
}
\label{fig:PBHspectrum}
\end{figure}

Finally, we define the mass function of PBHs as
\begin{equation}
   f(M_\mathrm{PBH})=\frac{1}{\left. \Omega_\mathrm{CDM}\right|_\mathrm{eq}} \frac{\mathrm{d}  \left. \Omega_\mathrm{PBH}\right|_\mathrm{eq}}{\mathrm{d}\ln M_\mathrm{PBH}},
\end{equation}
where $f(M_\mathrm{PBH})$ is normalized to integrate to unity if PBHs make up the entirety of dark matter. Expressing the integral in equation \eqref{eqn:beta} in terms of $M_\mathrm{PBH}$ using equation \eqref{eqn:CofM} allows us to write the final expression for PBH mass function
\begin{equation}
    f(M_\mathrm{PBH}) = \frac{M_\mathrm{PBH}}{\left. \Omega_\mathrm{CDM}\right|_\mathrm{eq}}\int\limits_{M_\mathrm{min}}^{M_\mathrm{max}} \mathrm{d}M_\mathrm{H}\left( \frac{M_\mathrm{eq}}{M_\mathrm{H}} \right)^{1/2}\frac{\mathrm{d}C_1(M_\mathrm{PBH})}{\mathrm{d}M_\mathrm{PBH}}\frac{M_\mathrm{PBH}}{M_\mathrm{H}}P(C_1(M_\mathrm{PBH}),M_\mathrm{H}).
\end{equation}
The PBH mass function $f(M_\mathrm{PBH})$ is shown in Figure \ref{fig:PBHspectrum}, for which PBHs make up the entirety of dark matter, and which evades constraints on the PBH abundance.

\section{Generation of tensor modes}
\label{5}

In this section, we compute the generation of tensor modes
$h_{ij}$ defined as
\begin{equation}
g_{ij}(t,\bm{x}) = a(t)^2\left(\delta_{ij}+\dfrac{1}{2}h_{ij}(t,\bm{x})\right)\ ,
\end{equation}
in our model.
There are apparently two significant contributions to the tensor mode. One is the perturbation of the two-form field itself, and the other is the curvature perturbation sourced by the two-form field. We will consider them in order.

\subsection{Primordial tensor modes}

We first calculate the primordial tensor power spectrum directly sourced by the two-form field.
The EoM for $h_{ij}$ is given by
\begin{equation}
\left(\partial_t^2 + 3H\partial_t - \nabla^2 \right)h_{ij} \simeq -\dfrac{4\bar{I}^2}{\Mpl^2a^4}\Pi^{lm}_{ij}\delta\dot{B}_{ln}\delta\dot{B}_{mn} \ ,
\end{equation}
where $\Pi^{lm}_{ij}$ is a projection operator into transverse and traceless components.
We decompose $h_{ij}$ into Fourier modes\footnote{Hereafter for tensor modes we will use the same notation of wave vector $\bm{k}$ as is used for scalar mode. But we notice that they are independent and hence are not related to each other.}
\begin{equation}
h_{ij} = \sum_{\lambda = +,\times}\int\dfrac{d\bm{k}}{(2\pi)^3}\hat{h}^\lambda_{\bm{k}}e^\lambda_{ij}(\hat{\bm{k}})e^{i\bm{k}\cdot\bm{x}} \ .
\end{equation}
% 
%%%%%%%%%%%%%%%%%%%%%%%%%%%%%%%%%%%%%%%%%%%
%%%%%%%%%%%%%%%%%%%%%%%%%%%%%%%%%%%%%
Using the relations, $\hat{h}^s_{\bm
{k}} = e^{s}_{ij}(\hat{\bm k})\hat{h}_{ij}(\bm k)$ and $\Pi_{ij}^{lm}e^s_{lm}(\hat{\bm k}) = e^s_{ij}(\hat{\bm k})$,
we obtain
\begin{align}
\left[ \partial_x^2 +1 - \dfrac{2}{x^2} \right](a\hat h^{s}_{\bm k}) = -e^{s}_{ij}(\hat{\bm{k}})\dfrac{4a^3}{k^2\Mpl^2} \int\dfrac{d\bm p}{(2\pi)^3}E_{\bm{p}}E_{\bm{k}-\bm{p}}\epsilon_{in}(\hat{\bm{p}})\epsilon_{jn}(\widehat{\bm{k}-\bm{p}})  \ . \label{eq: ten}
\end{align}
Regarding the calculation of the polarization tensors,
we use the following identities
\begin{align}
e^+_{ij}(\hat{\bm{k}})\epsilon_{in}(\hat{\bm{p}})\epsilon_{jn}(\widehat{\bm{k}-\bm{p}}) &= -\dfrac{1}{\sqrt{2}}\sin\theta_{\hat{\bm{p}}}\sin\theta_{\widehat{\bm{k}-\bm{p}}}\cos2\phi_{\hat{\bm{p}}} \ , \\
e^\times_{ij}(\hat{\bm{k}})\epsilon_{in}(\hat{\bm{p}})\epsilon_{jn}(\widehat{\bm{k}-\bm{p}}) &= -\dfrac{1}{\sqrt{2}}\sin\theta_{\hat{\bm{p}}}\sin\theta_{\widehat{\bm{k}-\bm{p}}}\sin2\phi_{\hat{\bm{p}}} \ .
\end{align}
Remarkably, they vanish in the integration over $\phi_{\hat{\bm p}}$.
Therefore, as has been shown in the previous studies \cite{Ohashi:2013qba,Obata:2018ilf}, the two-form field does not directly produce the primordial tensor modes at leading order\footnote{We could interpret this property as the fact that the spatial component of two-form field $B_{ij}$ does not have the spin-1 or spin-2 modes which are necessary to source tensor modes in view of spin conservation law. Although the non-dynamical component $B_{0i}$ has spin-1 modes \cite{Obata:2018ilf}, its contribution is sub-dominant in comparison with the induced tensor modes discussed in next section.}.

\subsection{Induced tensor modes}

Next, we compute the tensor mode induced by the scalar modes after inflation by following a method developed in the previous works~\cite{Saito:2009jt,Saito:2008jc,Inomata:2016rbd,Ando:2017veq,Ando:2018qdb,Espinosa:2018eve,Kohri:2018awv,Domenech:2019quo,Domenech:2020kqm} (see e.g. \cite{Domenech:2021ztg} for detailed review).
We take the conformal Newtonian gauge:
\begin{equation}
ds^2 = a(\tau)^2\left[ -(1+2\Phi)d\tau^2 + \left\{(1-2\Psi)\delta_{ij} + \dfrac{1}{2}h_{ij}\right\}dx^idx^j\right] \label{eq: re} \ ,
\end{equation}
where we neglected the vector perturbations
and assumed that the two scalar perturbations $\Phi$ and $\Psi$ satisfy the condition of no anisotropic pressure: $\Phi = \Psi$.
The density power spectrum has the peak at $k\sim 10^{13}\mathrm{Mpc}^{-1}$ in our setup, and such modes reenter the horizon during the radiation-dominated era, $\tau < \tau_{\rm eq}$.
Then, the equation of motion for tensor mode is given by
\begin{align}
&\left[\partial_\tau^2  - \nabla^2 \right](ah_{ij}) = -4a\Pi_{ij}^{lm}\mathcal{S}_{lm} \ , \\
&\mathcal{S}_{ij} \equiv 4\Psi\partial_i\partial_j\Psi + 2\partial_i\Psi\partial_j\Psi - \dfrac{1}{\mathcal{H}^2}\partial_i(\Psi' + \mathcal{H}\Psi)\partial_j(\Psi' + \mathcal{H}\Psi)
,
\end{align}
with $\mathcal H \equiv aH$,
and we obtain the solution of the induced tensor mode in the momentum space $h^s_{\bm k,\rm i}(\tau)$:
\begin{align}
h^s_{\bm k,\rm i}(\tau) &= \dfrac{4}{a(\tau)}\int_0^\infty d\tau' a(\tau')G_k(\tau, \tau')\mathcal{S}_{\bm k}(\tau') \ , \\
G_k(\tau, \tau') &\equiv \Theta(\tau - \tau')\dfrac{1}{k}\sin(k\tau-k\tau') \label{eq: indG} \ , \\
\mathcal{S}_{\bm k}(\tau) &= e_{ij}^{s}(\hat{\bm{k}})\int\dfrac{d\bm{p}p_ip_j}{(2\pi)^3}\left[ 3\Psi_{\bm{p}}\Psi_{\bm k - \bm{p}} + \dfrac{1}{\mathcal{H}}\left(\Psi_{\bm{p}}\Psi'_{\bm k - \bm{p}} + \Psi'_{\bm{p}}\Psi_{\bm k - \bm{p}} \right) + \dfrac{1}{\mathcal{H}^2}\Psi'_{\bm{p}}\Psi'_{\bm k - \bm{p}}  \right] \label{eq: Sk} \ .
\end{align}
To evaluate \eqref{eq: Sk}, we decompose $\Psi_{\bm k}(\tau)$ into the primordial field $\psi_{\bm k}$ and the transfer function $\Psi(k\tau)$ during the radiation-dominated era: 
\begin{align}
    \Psi_{\bm k}(\tau) &= \psi_{\bm k}\Psi(k\tau) \ , \\
    \Psi(k\tau) &= \dfrac{9}{(k\tau)^2}\left[ \dfrac{\sin(k\tau/\sqrt{3})}{k\tau/\sqrt{3}} - \cos(k\tau/\sqrt{3}) \right] \label{eq: tra} \ .
\end{align}
Using $\psi_{\bm k} = -2\zeta_{\bm k}/3$ at the radiation-dominated era, we can obtain the power spectrum of the tensor mode $\mathcal{P}_h(k,\tau)$ as 
\begin{align}
\sum_{\lambda=+,\times}\langle h^\lambda_{\bm k, \rm i}(\tau)h^{\lambda}_{\bm k', \rm i}(\tau) \rangle &= \dfrac{128}{81}\dfrac{1}{\tau^2k^3k'^3}\int\dfrac{d\bm{p}d\bm{q}}{(2\pi)^6}\sum_{\lambda=+,\times}e^\lambda_{ij}(\hat{\bm k})p_ip_je^{\lambda}_{kl}(\hat{\bm k}')q_kq_l \notag \\
&\times\mathcal{I}(p/k,|\bm k - \bm p|/k, k\tau)\mathcal{I}(q/k',|\bm k' - \bm q|/k', k'\tau)\langle \zeta_{\bm p}\zeta_{\bm k - \bm p}\zeta_{\bm q}\zeta_{\bm k' - \bm q} \rangle \notag \\
&\equiv (2\pi)^3\delta(\bm k+\bm k')\dfrac{2\pi^2}{k^3}\mathcal{P}_h(k,\tau) \ , \label{eq: ind}
\end{align}
where $\mathcal{I}$ is given by the time integral of transfer functions (its explicit form is given in Appendix \ref{app:calckernel}).

Then, we evaluate the current logarithmic energy density of GWs
 \begin{equation}
 \Omega_\mathrm{GW}(k) \equiv \dfrac{1}{\rho_c}\dfrac{d\rho_\mathrm{GW}}{d\ln k} \ ,
 \end{equation}
where $\rho_c = 3\Mpl^2H_0^2$ is the critical density of present universe.
In terms of the entropy conservation law, it is found to be ~\cite{Ando:2017veq}
\begin{align}
	\Omega_\mathrm{GW}(k)h^2
	&\simeq 
3.4\times10^{-7}\frac{\Omega_{r,0}h^2}{4.2\times10^{-5}}
	\left(\frac{\mathrm{g}_{*,c}}{106.75}\right)^{-1/3}
	\left(\dfrac{k}{aH(\tau_c)}\right)^2
	\overline{\mathcal{P}_h(k,\tau_c)} \ ,
\end{align}
where $\Omega_{r,0}$ is the radiation density parameter at present, $h \simeq 0.68$ is the dimensionless Hubble parameter, 
$\tau_c$ is the time when the production of GWs finished,
and $\mathrm{g}_{*,c}$ is the relativistic degree of freedom at $\tau_c$. 
The over line on $\mathcal{P}_h(k,\tau_c)$ denotes the average of the oscillation of $\mathcal{I}$ in \eqref{eq: ind}. 
The averaged product of two $\mathcal{I}$'s can be approximately decomposed as the following analytical formula:
\begin{align}
    \overline{
    \mathcal{I}(\tfrac{p}{k},\tfrac{|\bm k - \bm p|}{k}, k\tau_c)\mathcal{I}(\tfrac{q}{k'},\tfrac{|\bm k' - \bm q|}{k'}, k'\tau_c)} \simeq \sum_{\sigma =1,2}
    \frac{1}{2} 
    K_\sigma(\tfrac{p}{k},\tfrac{|\bm k - \bm p|}{k})
    K_\sigma(\tfrac{q}{k'},\tfrac{|\bm k' - \bm q|}{k'}) \ ,
\end{align}
where $K_i$ is obtained as \cite{Kohri:2018awv}
\begin{align}
K_1(\nu, u) &= \dfrac{27(\nu^2+u^2-3)}{16\nu^3u^3}\left( -4\nu u + (\nu^2 + u^2 - 3)\log\left|\dfrac{3-(\nu+u)^2}{3-(\nu - u)^2}\right|\right) \ , \label{eq: K1} \\
K_2(\nu, u) &= \dfrac{27\pi(\nu^2+u^2-3)^2}{16\nu^3u^3}\Theta(\nu + u - \sqrt{3}) \label{eq: K2}
\end{align}
(more explanations on how to obtain this formula is written in Appendix \ref{app:calckernel}).

\begin{figure}[thpb]    
\begin{center}     
\begin{tikzpicture} 
\begin{feynhand}    
    \vertex [particle] (i1) at (-7,0) {$h$};
    \vertex [particle] (f1) at (-3,0) {$h$};
    \vertex [dot] (w1) at (-6.3,0) {};
    \vertex [dot] (w2) at (-3.7,0) {};
    \vertex [dot] (w3) at (-5.7,0.7) {};
    \vertex [dot] (w4) at (-4.3,0.7) {};
    \vertex [dot] (w5) at (-5.7,-0.7) {};
    \vertex [dot] (w6) at (-4.3,-0.7) {};
    \propag [plain] (i1) to (w1);
    \propag [sca] (w1) to (w3);
    \propag [sca] (w1) to (w5);
     \propag [sca] (w2) to (w4);
    \propag [sca] (w2) to (w6);
     \propag [photon] (w3) to [in=135,out=45] (w4);
      \propag [photon] (w3) to [in=-135,out=-45] (w4);
       \propag [photon] (w5) to [in=135,out=45] (w6);
      \propag [photon] (w5) to [in=-135,out=-45] (w6);
    \propag [plain] (w2) to (f1);
    \vertex [particle] (i1) at (-2,0) {$h$};
    \vertex [particle] (f1) at (2,0) {$h$};
    \vertex [dot] (w1) at (-1.3,0) {};
    \vertex [dot] (w2) at (1.3,0) {};
    \vertex [dot] (w3) at (-0.7,0.7) {};
    \vertex [dot] (w4) at (0.7,0.7) {};
    \vertex [dot] (w5) at (-0.7,-0.7) {};
    \vertex [dot] (w6) at (0.7,-0.7) {};
    \propag [plain] (i1) to (w1);
    \propag [sca] (w1) to (w3);
    \propag [sca] (w1) to (w5);
     \propag [sca] (w2) to (w4);
    \propag [sca] (w2) to (w6);
     \propag [photon] (w3) to (w4);
      \propag [photon] (w3) to (w5);
       \propag [photon] (w5) to (w6);
      \propag [photon] (w4) to (w6);
    \propag [plain] (w2) to (f1);
     \vertex [particle] (i1) at (3,0) {$h$};
    \vertex [particle] (f1) at (7,0) {$h$};
    \vertex [dot] (w1) at (3.7,0) {};
    \vertex [dot] (w2) at (6.3,0) {};
    \vertex [dot] (w3) at (4.3,0.7) {};
    \vertex [dot] (w4) at (5.7,0.7) {};
    \vertex [dot] (w5) at (4.3,-0.7) {};
    \vertex [dot] (w6) at (5.7,-0.7) {};
    \propag [plain] (i1) to (w1);
    \propag [sca] (w1) to (w3);
    \propag [sca] (w1) to (w5);
     \propag [sca] (w2) to (w4);
    \propag [sca] (w2) to (w6);
     \propag [photon] (w3) to (w4);
      \propag [photon] (w3) to (w6);
       \propag [photon] (w5) to (w4);
      \propag [photon] (w5) to (w6);
    \propag [plain] (w2) to (f1);
    \end{feynhand}
\end{tikzpicture}
\caption{Diagrams of the contractions of 2-form field to calculate the power spectrum of induced GWs. We label these diagrams as ``Reducible" (left), ``Planar" (center) and ``Non-Planar" (right). The external solid line represents the gravitational wave perturbation $h^{+/\times}$. The intermediate dashed (wiggly) line represents the curvature perturbation $\zeta$ (the two-form field $B$).}
\label{fig:diagrams}
\end{center}
\end{figure}
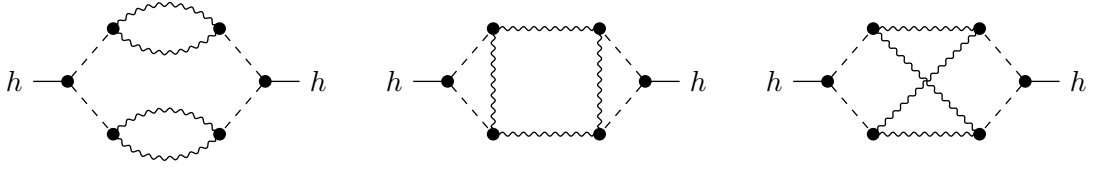

Now, let us compute the 4-point correlation function $\langle\zeta^4\rangle$ in \eqref{eq: ind}.
The explicit form of $\zeta$ with a momentum mode $\bm{k}$ is given by the sourced scalar mode \eqref{Greenfucntion} evaluated at the time of inflation end $\tau = \tau_{\rm end}$:
\begin{align}
\zeta_{\bm{k}} = -\dfrac{H}{\dot{\varphi}}\delta\varphi_{\bm{k}} = \dfrac{1}{H^2\Mpl^2\epsilon_H k}\int_{-\infty}^{\infty} \dfrac{d\tau'}{\tau'^3} (-\tau_{\rm end})G_R(\tau_{\rm end},\tau')n(\tau')\delta\rho_{E, \bm{k}} \ ,
\end{align}
where we have used \eqref{eq: n} and the slow-roll approximation.
It is important to notice that it cannot be merely replaced by $\langle\zeta^2\rangle^{2}$ since $\zeta$ is non-Gaussian: $\zeta \sim \delta B \delta B$.
In particular, we need to calculate the 8-point correlation function of primordial electric component of two-form field in $\langle\zeta^4\rangle$:
\begin{align}
&\langle\zeta_{\bm{p}}\zeta_{\bm{k}-\bm{p}}\zeta_{\bm{q}}\zeta_{\bm{k}'-\bm{q}} \rangle \notag \\
&=\left(\dfrac{1}{2H\Mpl\sqrt{\epsilon_H}}\right)^8\int \dfrac{d\bm{p}_1d\bm{p}_2d\bm{q}_1d\bm{q}_2}{(2\pi)^{12}}\prod_{i=1}^4\left[\int_{-\infty}^{+\infty} \dfrac{d\tau_i}{\tau_i^3}n(\tau_i)(-\tau_{\rm end})G_R(\tau_{\rm end},\tau_i)\right] \notag \\
&\times \epsilon_{ij}(\hat{\bm{p}}_1)\epsilon_{ij}(\widehat{\bm{p}-\bm{p}_1})\epsilon_{kl}(\hat{\bm{p}}_2)\epsilon_{kl}(\widehat{\bm{k}-\bm{p}-\bm{p}_2})\epsilon_{op}(\hat{\bm{q}}_1)\epsilon_{op}(\widehat{\bm{q}-\bm{q}_1})\epsilon_{qr}(\hat{\bm{q}}_2)\epsilon_{qr}(\widehat{\bm{k}'-\bm{q}-\bm{q}_2}) \notag \\
&\times\langle \hat{E}_{\bm{p}_1}(\tau_1)\hat{E}_{\bm{p}-\bm{p}_1}(\tau_1)\hat{E}_{\bm{p}_2}(\tau_2)\hat{E}_{\bm{k}-\bm{p}-\bm{p}_2}(\tau_2)\hat{E}_{\bm{q}_1}(\tau_3)\hat{E}_{\bm{q}-\bm{q}_1}(\tau_3)\hat{E}_{\bm{q}_2}(\tau_4)\hat{E}_{\bm{k}'-\bm{q}-\bm{q}_2}(\tau_4) \rangle \ . \label{eq: 4}
\end{align}
As shown in the previous studies \cite{Garcia-Bellido:2017aan,Cai:2018dig,Unal:2018yaa,Adshead:2021hnm}, these non-Gaussian sources contribute to the generation of induced GWs via three distinct diagrams shown in Figure~\ref{fig:diagrams}.
In the left diagram (called ``Reducible" diagram), the contraction is taken as $\langle \zeta_{\bm p}\zeta_{\bm k - \bm p}\zeta_{\bm q}\zeta_{\bm k' - \bm q} \rangle
 \to \braket{\zeta_{\bm p}\zeta_{\bm q}} \braket{\zeta_{\bm k - \bm p}\zeta_{\bm k' - \bm q}}$ and is therefore simply replaced by the sourced power spectrum of curvature perturbation $\mathcal{P}_\zeta$.
To perform this, let us calculate the contribution to the reducible diagram.
Using the slow-roll approximation, it leads to
\begin{align}
\eqref{eq: 4}|_{\rm reducible} &=8\left(\dfrac{1}{2H\Mpl\sqrt{\epsilon_H}}\right)^8(2\pi)^6\delta(\bm{k}+\bm{k}')\delta(\bm{p}+\bm{q})\int \dfrac{d\bm{p}_1d\bm{p}_2}{(2\pi)^6}\prod_{i=1}^4\left[\int_{-\infty}^{\tau} \dfrac{d\tau_i}{\tau_i}n(\tau_i)\dfrac{\tau_i^3-\tau^3}{3\tau_i^3}\right]\epsilon_r \notag \\
&\times E_{\bm{p}_1}(\tau_1)E^*_{\bm{p}_1}(\tau_3)E_{\bm{p}-\bm{p}_1}(\tau_1)E^*_{\bm{p}-\bm{p}_1}(\tau_3)E_{\bm{p}_2}(\tau_2)E^*_{\bm{p}_2}(\tau_4)E_{\bm{k}-\bm{p}-\bm{p}_2}(\tau_2)E^*_{\bm{k}-\bm{p}-\bm{p}_2}(\tau_4) \ , \label{eq: redu}
\end{align}
%
%///////////////////////////////////////////////////////////////////////////////////%
%///////////////////////////////////////////////////////////////////////////////////%
%
where we have used $G_R(\tau,\tau_i) \equiv \Theta(\tau-\tau_i)(\tau^3-\tau_i^3)/(3\tau\tau_i)$ and defined
\begin{align}
\epsilon_r(\bm{k}, \bm{p}, \bm{p}_1, \bm{p}_2) &\equiv \epsilon_{ij}(\hat{\bm{p}}_1)\epsilon_{ij}(\widehat{\bm{p}-\bm{p}_1})\epsilon_{kl}(\hat{\bm{p}}_2)\epsilon_{kl}(\widehat{\bm{k}-\bm{p}-\bm{p}_2}) \notag \\
&\times \epsilon_{op}(-\hat{\bm{p}_1})\epsilon_{op}(-\widehat{(\bm{p}-\bm{p}_1)})\epsilon_{qr}(-\hat{\bm{p}_2})\epsilon_{qr}(-\widehat{(\bm{k}-\bm{p}-\bm{p}_2)}) \ .
\end{align}
The factor $8$ comes from the symmetry.
Therefore, we can rewrite \eqref{eq: redu} as the simplest form
\begin{equation}
\eqref{eq: redu} = 8(2\pi)^6\delta(\bm{k}+\bm{k}')\delta(\bm{p}+\bm{q})\dfrac{2\pi^2}{p^3}\dfrac{2\pi^2}{|\bm{k}-\bm{p}|^3}\mathcal{P}_\zeta(\bm{p})\mathcal{P}_\zeta(\bm{k}-\bm{p}) \ .
\label{eq_reducible_simple}
\end{equation}
This is the same expression of the power spectrum of induced GWs as that from the standard scenario in which the curvature perturbation is Gaussian~\cite{Saito:2009jt,Saito:2008jc,Inomata:2016rbd,Ando:2017veq,Ando:2018qdb,Espinosa:2018eve,Kohri:2018awv,Domenech:2019quo,Domenech:2020kqm}. However, we notice that the relationship of abundance between PBH mass function and the curvature power spectrum is different because of the non-Gaussianity of $\zeta$ in our scenario.
In the other two contributions (``Planar" and ``Non-Planar" diagrams),
$\delta B$s are contracted across more than two $\zeta$s, and hence we cannot replace the contracted $\zeta^2$ with $\mathcal{P}_\zeta$.
The contractions of two-form fields in Planar diagram lead to
\begin{align}
&\eqref{eq: 4}|_{\rm planar} = 32\left(\dfrac{1}{2H\Mpl\sqrt{\epsilon_H}}\right)^8(2\pi)^3\delta(\bm{k}+\bm{k}')\int \dfrac{d\bm{p}_1}{(2\pi)^3}\prod_{i=1}^4\left[\int_{-\infty}^{\tau} \dfrac{d\tau_i}{\tau_i}n(\tau_i)\dfrac{\tau_i^3-\tau^3}{3\tau_i^3}\right]\epsilon_p \notag \\
&\times E_{p_1}(\tau_1)E^*_{p_1}(\tau_2)E_{|\bm{p}-\bm{p}_1|}(\tau_1)E^*_{|\bm{p}-\bm{p}_1|}(\tau_3)E_{|\bm{p}+\bm{q}-\bm{p}_1|}(\tau_3)E^*_{|\bm{p}+\bm{q}-\bm{p}_1|}(\tau_4)E_{|\bm{k}-\bm{p}+\bm{p}_1|}(\tau_2)E^*_{|\bm{k}-\bm{p}+\bm{p}_1|}(\tau_4) \ , \label{eq:planar1}
\end{align}
where
\begin{align}
\epsilon_p(\bm{k}, \bm{p}, \bm{q}, \bm{p}_1) &\equiv \epsilon_{ij}(\hat{\bm{p}}_1)\epsilon_{ij}(\widehat{\bm{p}-\bm{p}_1})\epsilon_{kl}(-\hat{\bm{p}}_1)\epsilon_{kl}(\widehat{\bm{k}-\bm{p}+\bm{p}_1}) \notag \\
&\times \epsilon_{op}(\widehat{\bm{p}+\bm{q}-\bm{p}_1})\epsilon_{op}(-\widehat{(\bm{p}-\bm{p}_1)})\epsilon_{qr}(-\widehat{(\bm{p}+\bm{q}-\bm{p}_1)})\epsilon_{qr}(-\widehat{(\bm{k}-\bm{p}+\bm{p}_1)}) \ . 
\end{align}
And that in Non-Planar diagram is given by
\begin{align}
&\eqref{eq: 4}|_{\rm non-planar} \notag \\
&=32\left(\dfrac{1}{2H\Mpl\sqrt{\epsilon_H}}\right)^8(2\pi)^3\delta(\bm{k}+\bm{k}')\int \dfrac{d\bm{p}_1}{(2\pi)^3}\prod_{i=1}^4\left[\int_{-\infty}^{\tau} \dfrac{d\tau_i}{\tau_i}n(\tau_i)\dfrac{\tau_i^3-\tau^3}{3\tau_i^3}\right]\epsilon_{n} \notag \\
&\times E_{\bm{p}_1}(\tau_1)E^*_{\bm{p}_1}(\tau_3)E_{\bm{p}-\bm{p}_1}(\tau_1)E^*_{\bm{p}-\bm{p}_1}(\tau_4)E_{\bm{k}-\bm{p}+\bm{q}+\bm{p}_1}(\tau_2)E^*_{\bm{k}-\bm{p}+\bm{q}+\bm{p}_1}(\tau_4)E_{-\bm{q}-\bm{p}_1}(\tau_2)E^*_{-\bm{q}-\bm{p}_1}(\tau_3) \ ,
\end{align}
where
\begin{align}
\epsilon_{n}(\bm{k}, \bm{p}, \bm{q}, \bm{p}_1) &\equiv \epsilon_{ij}(\hat{\bm{p}}_1)\epsilon_{ij}(\widehat{\bm{p}-\bm{p}_1})\epsilon_{kl}(\widehat{\bm{k}-\bm{p}+\bm{q}+\bm{p}_1})\epsilon_{kl}(-\widehat{(\bm{q}+\bm{p}_1)}) \notag \\
&\times \epsilon_{op}(-\hat{\bm{p}}_1)\epsilon_{op}(\widehat{\bm{q}+\bm{p}_1})\epsilon_{qr}(-\widehat{(\bm{k}-\bm{p}+\bm{q}+\bm{p}_1)})\epsilon_{qr}(-\widehat{(\bm{p}-\bm{p}_1)}) \ .
\end{align}
These contributions have been computed numerically in the previous work in the case of not two-form field but U(1) gauge field~\cite{Garcia-Bellido:2017aan} or the local-type non-Gaussian field \cite{Adshead:2021hnm} and it was found that the Reducible diagram and Planar diagram had the same order of contributions to the spectrum, whereas that from Non-Planar diagram is suppressed compared to the other two diagrams\footnote{
More precisely, the Reducible diagram is dominant on the high momentum side of the peak, whereas the Planar diagram becomes comparable to the Reducible diagram at the tail of the low momentum side.
}.
Therefore, in this work, we numerically estimate the contributions only from the Reducible diagram and the Planar diagram.

The computation of the Reducible diagram is easily performed since we can factorize the two loops of $ B$-$ B$ into the power spectrum of $\zeta$.
The formula of reducible diagram is written in the same way to that of induced GWs without non-Gaussianity~\cite{Saito:2009jt,Saito:2008jc,Inomata:2016rbd,Ando:2017veq,Ando:2018qdb,Espinosa:2018eve,Kohri:2018awv,Domenech:2019quo,Domenech:2020kqm}
\begin{align}
\Omega_\mathrm{GW}(k)h^2
&=
    0.39
	\left(\frac{\mathrm{g}_{*,c}  }{106.75  }\right)^{-1/3}
	\Omega_{r,0}h^2 
    \frac{8}{243}
	\int^\infty_0\mathrm{d} \nu
	\int^{1+\nu}_{ \left|1-\nu\right| }\df u
	 \mathcal P_\zeta(ku)\mathcal P_\zeta(k\nu)
	\frac{\nu^2}{u^2}
\nonumber
\\&	\qquad\times
    \left(  1- \frac{(1+\nu^2-u^2)^2}{4\nu^2}  \right)^2
	\overline{\mathcal I(u,\nu,k\tau_c)}^2
	\label{eq_Omega_GW_Reducible}
      ,
\end{align}
where we followed expressions in \cite{Kawasaki:2019hvt}.
On the other hand, the Planar diagram has three loops of momentum integrals which cannot be fuctorized into $\mathcal{P}_\zeta$, so that the simple analytical method cannot be used.
In this work, we investigate an efficient method to calculate the Planar type diagram and show the reduced formulae on the Planar-type GWs in Appendix.~\ref{sec_calc_gws}.
The basic idea is as follows: the Planar diagram has two $\zeta$-$\zeta$-$ B$ loops and we can firstly contract them as two effective vertices of $h$-$ B$-$ B$.
Afterwards, we can calculate only one residual $ B$-$ B$ loop in terms of the obtained vertices.
To reduce the computational costs, we approximate the profile of two-form mode function as a Gaussian fitting function \eqref{eq: fitting}.

\begin{figure}[tbhp]
    \begin{center}
    \includegraphics[width=1\textwidth]{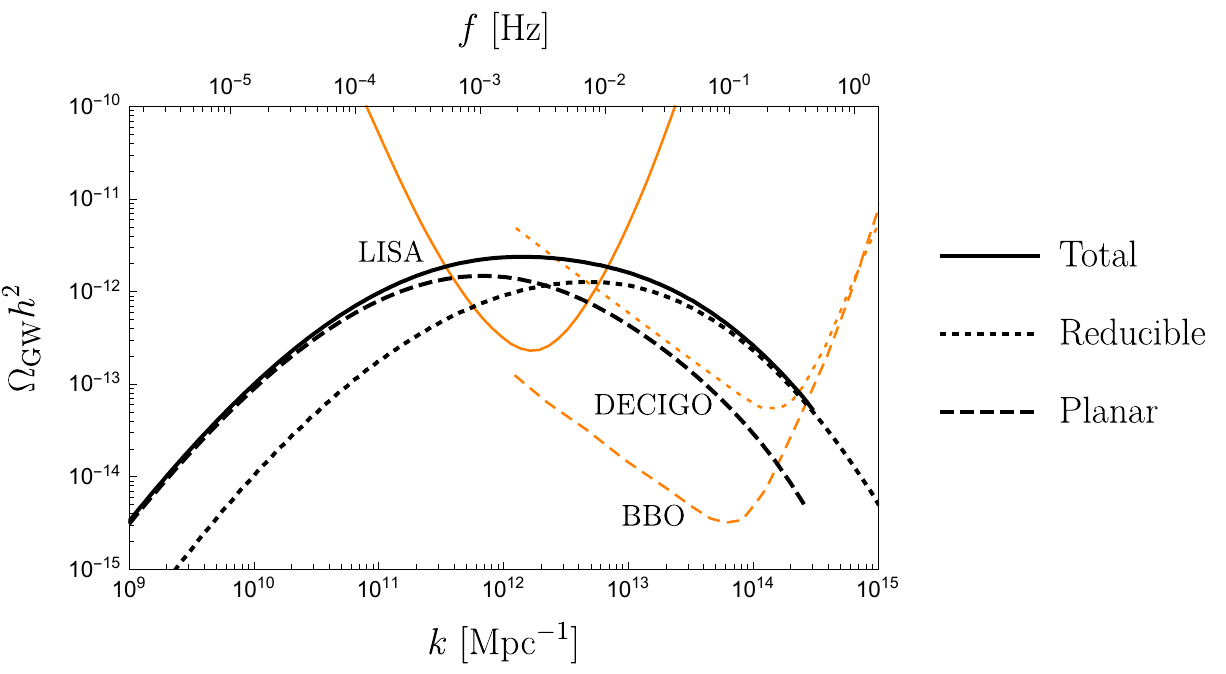}
    \end{center}
     \caption{
     The power spectrum of induced GWs is shown.
     We evaluate the Reducible-type GWs (dotted black line) in \eqref{eq_Omega_GW_Reducible}, the Planar-type GWs (dashed black line) in \eqref{eq_Omega_GWs_planar}, and a sum of them (solid black line).
      We also show the future sensitivity curves of interferometers as orange lines; LISA (solid)~\cite{Smith:2019wny}, DECIGO (dotted), and BBO (dashed), respectively~\cite{Moore:2014lga}.
        }
      \label{fig_inducedGW}
\end{figure}

We numerically estimate $\Omega_\mathrm{GW}(k)$ for Reducible diagram \eqref{eq_Omega_GW_Reducible} (dotted black line) and Planar diagram \eqref{eq_Omega_GWs_planar} (dashed black line) shown in Fig.~\ref{fig_inducedGW}.
Both the Reducible and the Planar type GWs have similar contributions, while the Planar type has a peak at smaller frequency than Reducible type.
This relation is consistent with the previous study on the axion-$U(1)$ inflationary model ~\cite{Garcia-Bellido:2017aan}\footnote{
The authors in \cite{Garcia-Bellido:2017aan} used a different computational method from our work. To simplify the procedure, they approximated the profile of the gauge mode function as a Dirac delta function peaked at the momentum scale where particle production mostly occurs (so called ``zero-width'' approximation), and reduced a number of integration variables to solve.}.
The GWs around this frequency is potentially observable with the future interferometer experiments like LISA, DECIGO, and BBO,
while the strength of GWs depends on the uncertainty on the PBH formation rate and the momentum dependence of the factor $\mathcal F$~\eqref{eq: F}.

%%%%%%%%%%%%%%%%%%%%%%%%%%%%%%%%%%%%%%%%%%%
%%%%%%%%%%%%%%%%%%%%%%%%%%%%%%%%%%%%%

%%%%%%%%%%%%%%%%%%%%%%%%%%%%%%%%%%%%%%%%%%%%
\section{Conclusion}
\label{6}
%%%%%%%%%%%%%%%%%%%%%%%%%%%%%%%%%%%%%%%%%%%%

In this work, we have suggested an inflationary model where the inflaton is kinetically coupled to a two-form field, and developed a possibility of particle production of the two-form field on intermediate scales during inflation.
Depending on the configuration of the kinetic function, the two-form field is enhanced at a later stage of inflation and amplifies the coupled curvature perturbation at second-order level.
Then, we showed that the sourced curvature perturbation can explain the formation of PBHs in sufficient abundance to act as dark matter.
This production mechanism would be similar to that from the model of $U(1)$ gauge field coupled with a dilatonic field \cite{Kawasaki:2019hvt} or an axionic field \cite{Garcia-Bellido:2016dkw,Garcia-Bellido:2017aan}.
Regarding tensor modes, however, the two-form field does not directly produce primordial tensor modes because of its antisymmetric feature\footnote{Although the integration of non-dynamical component of two-form field may provide the primordial tensor modes, its contribution would be sub-dominant. Also, it does not provide any chirality in tensor mode like the model of axion-$U(1)$ coupling \cite{Garcia-Bellido:2016dkw,Garcia-Bellido:2017aan}.}.
Therefore, the absence of primordial GWs could be a unique prediction from two-form field model.
On the other hand, our scenario predicts the associated induced GWs after inflation.
Since the primordial curvature perturbation is sourced by the second-order of two-form field, three diagrammatic contributions appear in the spectrum of induced GWs: Reducible, Planar, and Non-Planar diagram.
We computed the contributions from Reducible and Planar diagrams and found that they prodive a comparable amount of induced tensor spectrum with different frequency regime, which modifies the relationship of spectral peaks between PBH mass function and induced GWs from the standard prediction in Gaussian field model.
We also showed that the resultant spectraum is potentially testable with future laser interferometers.

It is interesting to consider the possibility that this model can also explain a recent NanoGrav event \cite{Arzoumanian:2020vkk}.
In our setup, however, the generation of sourced curvature perturbation on that scale would be challenging.
The reason is as follows.
To provide NanoGrav event, we would need the particle production of two-form field occurring on scales much greater than scales corresponding to a few number of e-foldings before inflation ends.
However, on such scales the slow-roll condition is sufficiently met and hence the value of $n$ (given by the speed of scalar field \eqref{eq: n}) does not drastically change in time. 
This feature would make it hard to provide the peaky spectral shape in PBH mass spectrum and associated induced GWs testable with NanoGrav event.
Therefore, we would need to extend this model to realize it and we leave further consideration of this to future study.

\section*{Acknowledgement}

This work was supported by JSPS KAKENHI Grant No. JP18K13537(T.F.), JP19J21974 (H.N.) and JP20H05859 (I.O.).
H.N. is also supported by the Advanced Leading Graduate Course for Photon Science.
I.O. acknowledges the support from JSPS Overseas Research Fellowship.
S.Y. was supported by a Humboldt Research Fellowship, and an MCSA Postdoctoral Fellowship.
This project has received funding from the European Union's Horizon 2020 research and innovation programme under the Marie Sk\l odowska-Curie grant agreement No 101029832.

\appendix

%///////////////////////////////////////////////////////////////////////////////////%
\section{Typical perturbation profile shapes}
\label{app:profiles}
%///////////////////////////////////////////////////////////////////////////////////%

In this section, we will describe a simple calculation of the relation between $\chi(r_m)$ and $\chi_{2,r}$ given in section \ref{4}. A more detailed and rigorous discussion of the probability density functions and correlation functions in the context of peaks theory can be found in reference \cite{Young:2022phe}.

For power spectra which have a relatively narrow peak, we can approximate that large, rare curvature perturbations $\zeta$ which form PBHs to be spherically symmetric and have a profile shape given generically by \cite{Bardeen:1985tr,Young:2019osy}
\begin{equation}
    \zeta = \chi^2(r) = C \left(\frac{\sin(k_* r)}{k_* r}\right)^2,
\end{equation}
where $C$ determines the amplitude of the perturbation and $k_*$ determines the scale of the perturbation. 

The characteristic scale of a perturbation $r_m$ is given by the scale at which the compaction function peaks, which can be found by solving \cite{Musco:2018rwt}
\begin{equation}
    \zeta'(r_m)+r_m\zeta''(r_m)=0,
\end{equation}
where a prime again denotes a derivative with respect to the radial coordinate $r$. For the profile shape given above, we can numerically calculate $r_m$, and the ratio between $\chi(r_m)$ and $\chi_{2,r}$ is then given, independently of $C$ and $k_*$, by
\begin{equation}
    \chi(r_m)/\chi_{2,r}\approx -0.756,
\end{equation}
which is the value used in section \ref{4}.

%///////////////////////////////////////////////////////////////////////////////////%
\section{Calculation of integral $\mathcal{I}$}
\label{app:calckernel}
%///////////////////////////////////////////////////////////////////////////////////%

In this appendix, we briefly present the computation of integral $\mathcal{I}$ in the power spectrum of induced GWs.
In terms of \eqref{eq: indG}-\eqref{eq: tra}, the original expression of $\mathcal{I}$ is obtained as
\begin{align}
\mathcal{I}(\nu, u, x) &= \int_0^xdy y\sin(x-y)\left[ 3\Psi(\nu y)\Psi(u y) + y\{ \Psi(\nu y)u\dfrac{d\Psi(u y)}{d(uy)} + \nu\dfrac{d\Psi(\nu y)}{d(\nu y)}\Psi(u y) \} \right. \notag \\
&\left. + y^2u \nu \dfrac{d\Psi(\nu y)}{d(\nu y)}\dfrac{d\Psi(u y)}{d(uy)} \right] \ .
\end{align}
Since $\sin(x-y) = \sin x \cos y - \cos x \sin y$, we can rewrite $\mathcal{I}$ as 
\begin{align}
\mathcal{I}(\nu, u, k\tau_c) 
&= 
\sin(k\tau_c) K_1(\nu,u,k\tau_c)
	-\cos(k\tau_c) K_2(\nu,u,k\tau_c) \ ,
	\\
	K_1(\nu,u,k\tau_c)
	&=
	\int^{k\tau_c}\mathrm{d} z
	z\cos(z)
	f(\nu k,uk,z/k) \ ,
	\\
	K_2(\nu,u,k\tau_c)
	&=
	\int^{k\tau_c}\mathrm{d} z
	z\sin(z)
	f(\nu k,uk,z/k) \ ,
	\\
	f(\nu k,uk,z/k)&=
	\left[
	3\Psi(\nu z)\Psi(uz)
	+\left(   \nu z\dot \Psi(\nu z)+\Psi(\nu z)   \right)
	\left(   uz\dot \Psi(uz)+\Psi(uz)   \right)
	\right] \ .
\end{align}
$\mathcal{I}(\nu, u, k\tau_c) $ consists of two oscillating terms with $\sin(k\tau_c)$ and $\cos(k\tau_c)$.

Since the GW spectrum observed today is sufficiently within the sub-horizon regime, we consider the late-time limit of $\mathcal{I}$: $k\tau_c\rightarrow\infty$. In this limit, $K_1(\nu,u,k\tau_c)$ and $K_2(\nu,u,k\tau_c)$ converge to constant and their asymptotic expressions \eqref{eq: K1} and \eqref{eq: K2} have been found \cite{Kohri:2018awv}.
Then, we need to take an oscillation average with respect to $k\tau_c$ to evaluate the amplitude of spectrum.
The product of two $\mathcal{I}$'s can be decomposed as
\begin{align}
    \overline{
    \mathcal{I}(\tfrac{p}{k},\tfrac{|\bm k - \bm p|}{k}, k\tau)\mathcal{I}(\tfrac{q}{k'},\tfrac{|\bm k' - \bm q|}{k'}, k'\tau)} =\sum_{\sigma =1,2}
    \frac{1}{2} 
    K_\sigma(\tfrac{p}{k},\tfrac{|\bm k - \bm p|}{k})
    K_\sigma(\tfrac{q}{k'},\tfrac{|\bm k' - \bm q|}{k'})
\end{align}
where we use the momentum conservation, $k=k'$, and $\overline{\sin(k\tau_c)^2}=\overline{\cos(k\tau_c)^2}=1/2$ and $\overline{\sin(k\tau_c)\cos(k\tau_c)}=0$, 
where overlines represent the average over $\tau_c$.

%///////////////////////////////////////////////////////////////////////////////////%
\section{Calculation of the Planar diagram of induced GWs}
\label{sec_calc_gws}
%///////////////////////////////////////////////////////////////////////////////////%

In this appendix, we present the computation of Planar diagram of induced GWs (see Figure \ref{fig:diagrams}).
By using the fitting function of $E_k$ \eqref{eq: fitting} and the following relationship 
\begin{align}
    \sum_{ij}\epsilon_{ij}(\hat{\bm{p}})\epsilon_{ij}(\hat{\bm{q}})
	 =  -2\frac{\bm{p}\cdot \bm{q}}{pq} \ ,
\end{align}
\eqref{eq:planar1} is rewritten as 
\begin{align}
\eqref{eq:planar1} &=
	32
	\left( \frac{H^2\mathcal F}{ 12\Mpl^2 \epsilon_H }\right)^4
	(2\pi)^3\delta^3(\bm k+\bm k') k^{-9}
	\int \frac{\df^3 r_1}{(2\pi)^3}  k^9 \notag
	\\&	\times
	\frac{E_{\text{peak}}(r_1)^2}{2r_1^3}
	\frac{E_{\text{peak}}(|\bm r_{pq} - \bm r_1 |)^2}{2|\bm r_{pq} - \bm r_1 |^3}
	\frac{E_{\text{peak}}(|\bm p - \bm r_1|)^2}{2|\bm p - \bm r_1|^3}
	\frac{E_{\text{peak}}(|\bm k-\bm p +\bm r_1|)^2}{2|\bm k-\bm p +\bm r_1|^3} \notag
	\\&	\times
	\frac{2\bm{r_1}\cdot (\bm{p}-\bm{r_1}) }{r_1|\bm{p}-\bm{r_1}|}
	\frac{2\bm{r_1}\cdot (\bm{k}-\bm{p}+\bm{r_1}) }{r_1|\bm{k}-\bm{p}+\bm{r_1}|}
	\frac{2(  \bm r_{pq} -\bm r_1 )\cdot ( \bm r_1- \bm p ) }{|  \bm r_{pq} -\bm r_1 ||\bm{r_1}-\bm{p}|}
	\frac{2(\bm r_{pq} - \bm r_1)\cdot (\bm{p}-\bm{k} - \bm r_1) }{|\bm r_{pq} - \bm r_1||\bm{p}-\bm{k} - \bm r_1|}
	,
\end{align}
where $\bm{r}_{pq} \equiv \bm{p} + \bm{q}$ and $\mathcal F$ (defined in \eqref{eq: F}) contains the integration over $\tau$.

Here, we investigate the efficient method to calculate the planar diagram.
Although all diagrams contain three loops and require $3\times3=9$ integration of momenta, the Reducible diagram is easy to estimate because we can firstly perform the momentum loop of two-form field.
Similarly, integration of triangle loops $\zeta$-$\zeta$-$ B$ in Planar diagram can be performed at first to derive the ``reduced vertex'' of $hBB$:

\begin{align}
	\Omega_\mathrm{GW}^\mathrm{planar}(k)h^2
&=
	A_\Omega  h^2
	\int \frac{\df^3 \bm r}{(2\pi)^3} 
	\sum_{s=+,\times}  
	\sum_{\sigma=1,2} 
	\frac{E_{\text{peak}}(| \bm r|)^2}{|\bm r|^3}
	\frac{E_{\text{peak}}(|\bm k-\bm r|)^2}{|\bm k-\bm r|^3}
	|F_{s,\sigma}(\bm r ,\bm k)|^2 , \\
	A_\Omega  h^2 &\equiv 
	0.39
	\left(\frac{\mathrm{g}_{*,c}  }{106.75  }\right)^{-1/3}
	\Omega_{r,0}h^2
	\frac{8}{243}
	\frac{1}{2}
	\left( \frac{H^2\mathcal F}{ 12\Mpl^2 \epsilon_H }\right)^4
	\frac{ 32 }{2\pi^2}
	\label{eq_Omega_GWs_planar}
\end{align}
where the reduced vertex $F_{s,\sigma}(\bm r, \bm k)$ is given by
\begin{align}
	&F_{s,\sigma}(\bm r, \bm k)
	\equiv
	\int  \frac{\df \bm{p}}{(2\pi)^3}
	e^{s}_{ij}(k)p_jp_j 
	K_\sigma (\tfrac{p}{k},\tfrac{|\bm k - \bm p|}{k})
	\frac{E_{\text{peak}}(|\bm p- \bm r |)^2}{|\bm p -\bm r|^3}
		\\&\quad\quad\times \nonumber
    \frac{	(\bm p - \bm r)\cdot	{\bm r}}{|\bm p - \bm r|	 |\bm r|}
    \frac{ (\bm p - \bm r )\cdot (\bm k- \bm{r} )}{|\bm p - \bm r ||\bm k- \bm{r} |}.
\end{align}
The numerical factor $A_\Omega$ is estimated as
\begin{align}
	A_\Omega  h^2 =
	4.15\times 10^{-43}~~ 
	\left(\frac{\mathrm{g}_{*,c}  }{106.75  }\right)^{-1/3}
	\frac{\Omega_{r,0}h^2}{4.2\times 10^{-5}}
	\left( \frac{ \mathcal{P}_{\zeta,v}(k_p) }{ 1.5\times 10^{-10} }\right)^4
	\mathcal F^4
	,
\end{align}
where $\mathcal{P}_{\zeta,v}(k_p)$ is a power spectrum of curvature fluctuations directly induced by vacuum mode fluctuations on the peak scale $k_p$, which is slightly larger than that in CMB scale
for the Starobinsky inflation model ~\eqref{eq:starobinsky_model_potential}.
The time integration $\mathcal F$ 
is $1\sim 1.1$ in our parameter setup.

\bibliographystyle{apsrev4-1}
\bibliography{Ref.bib}

\end{document}